\documentclass[12pt]{article}
\pdfoutput=1
\usepackage{amssymb}
\usepackage{amsmath}
\usepackage{latexsym}
\usepackage[usenames]{color}
\usepackage{fancybox}
\usepackage{simplewick}
\usepackage{comment}
\usepackage{cite}
\usepackage{framed}
\usepackage{booktabs}                       
\usepackage{multirow,bigdelim}              
\usepackage{longtable}                      
\definecolor{shadecolor}{rgb}{0.9,0.9,0.95}
\usepackage{setspace}
\usepackage{graphicx} 
\usepackage[usenames,dvipsnames]{xcolor}
\usepackage[setpagesize=false,pagebackref=false, linktocpage, bookmarksopen=true, colorlinks=true, linkcolor=RoyalBlue,citecolor=Maroon,urlcolor=Maroon]{hyperref}

\def\beq{\begin{equation}}
\def\eeq{\end{equation}}
\def\pmatrix#1#2{\left( 
\begin{array}{#1}
#2\end{array} 
\right)}
\def\del {\partial}

\def \S {\mathbf{S}}

\def\comma{\,,}
\def\period{\,.}

\def\XXint#1#2#3{{\setbox0=\hbox{$#1{#2#3}{\int}$}
     \vcenter{\hbox{$#2#3$}}\kern-.5\wd0}}

\begin{document}
\thispagestyle{empty}

\renewcommand{\thefootnote}{\fnsymbol{footnote}}
\setcounter{page}{1}
\setcounter{footnote}{0}
\setcounter{figure}{0}
\begin{flushright}
{\tt CERN-TH-2021-175}
\end{flushright}
\vspace{1cm}
\begin{center}
{\large \bf
Crosscap States in Integrable Field Theories and Spin Chains
\par}

\vspace{1cm}

\textrm{Jo\~{a}o Caetano,  Shota Komatsu}
\\ \vspace{1cm}
\footnotesize{\textit{
Department of Theoretical Physics, CERN, 1211 Meyrin, Switzerland
}  
\vspace{0.5cm}
}

{\tt  joao.dias.caetano.silva$\otimes$cern.ch, shota.komatsu$\otimes$cern.ch}

\par\vspace{2cm}

\textbf{Abstract}\vspace{2mm}
\end{center}
\noindent
We study crosscap states in integrable field theories and spin chains in $1+1$ dimensions. We derive an exact formula for overlaps between the crosscap state and any excited state in integrable field theories with diagonal scattering. We then compute the {\it crosscap entropy}, i.e.~the overlap for the ground state, in some examples. In the examples we analyzed, the result turns out to decrease monotonically along the renormalization group flow except in cases where the discrete symmetry is spontaneously broken in the infrared. 
We next introduce crosscap states in integrable spin chains, and obtain determinant expressions for the overlaps with energy eigenstates. These states are long-range entangled and their entanglement entropy grows linearly until the size of the subregion reaches half the system size. This property is reminiscent of pure-state black holes in holography and makes them interesting for use as initial conditions of quantum quench. As side results, we propose a generalization of Zamolodchikov's staircase model to flows between $D$-series minimal models, and discuss the relation to fermionic minimal models and the GSO projection.

\setcounter{page}{1}
\renewcommand{\thefootnote}{\arabic{footnote}}
\setcounter{footnote}{0}
\setcounter{tocdepth}{2}
\numberwithin{equation}{section}
\newpage
\tableofcontents
\parskip 5pt plus 1pt   \jot = 1.5ex
\newpage
\section{Introduction\label{sec:intro}}
Boundaries and defects enrich the study of quantum field theory (QFT) and statistical mechanics. In QFT, they give us access to the dynamical information of the theory that cannot be studied by local operators. A notable example is the Wilson-'t Hooft loop in gauge theories, which is the order parameter for confinement. In statistical mechanics on the other hand, there is a whole zoo of critical phenomena associated with boundaries and defects. One phenomenon that has attracted much attention in the past is the Kondo effect \cite{kondo1964resistance} i.e.~the anomalous behavior of the electrical resistivity due to magnetic impurities.

In $1+1$ dimensions, there are several ways to study non-perturbative dynamics of boundaries and defects. A  popular approach is to focus on fixed points of the renormalization group (RG) flow and use the techniques of two-dimensional conformal field theory (CFT). For example, in \cite{Ishibashi:1988kg,Cardy:1989ir,Cardy:1991tv} a systematic construction of conformal boundary conditions was discussed, and in \cite{Behrend:1998fd,Behrend:1999bn} a complete classification for the minimal models was obtained. Another approach is to study {\it integrable field theories}---theories with infinitely many conserved charges that are amenable to analytical calculations. In these theories, one can follow the dynamics of special boundaries, called integrable boundaries, all the way from UV to IR along the RG flow \cite{Ghoshal:1993tm,Dorey:1999cj,Dorey:2009vg,Dorey:2004xk,Dorey:2010ub}. Both approaches were successfully applied to the Kondo problem \cite{Affleck:1990by,Affleck:1990iv,Andrei:1980fv,wiegmann1981exact}.

Recently, the boundaries in $1+1$ dimensions have attracted renewed interest in the context of quantum quench \cite{Calabrese:2006rx,Calabrese:2007mtj,Caux:2013ra,Caux:2016esd}. The states produced by the boundaries, known as the {\it boundary states}, have often been used as initial conditions for quench protocols \cite{Calabrese:2006rx}. Since they are defined locally by the boundary condition, the boundary states are short-range entangled and provide ideal initial states for studying the growth of entanglement \cite{Calabrese:2007mtj}. Analogs of integrable boundary states are known also in spin chains (see e.g.~\cite{brockmann2014gaudin,brockmann2014neel,pozsgay2014overlaps,Foda:2015nfk,Pozsgay:2018ixm,Piroli:2018ksf,Pozsgay:2018dzs,DeLeeuw:2019ohp}). Taking them as initial conditions, precision computations of the quench dynamics in lattice systems were performed in the literature \cite{Caux:2013ra,Caux:2016esd,Piroli:2018amn}.

Moreover, boundaries in integrable systems play an important role in quite a different discipline, namely in the study of the AdS/CFT correspondence \cite{Maldacena:1997re}. A prototypical example of the AdS/CFT correspondence is the duality between $\mathcal{N}=4$ supersymmetric Yang-Mills theory (SYM) in four dimensions and type IIB string theory on AdS$_5\times $S$^5$ spacetime. Recent works have shown that various observables in $\mathcal{N}=4$ SYM, such as one-point functions in defect CFT \cite{deLeeuw:2015hxa,Buhl-Mortensen:2015gfd,Buhl-Mortensen:2017ind,Komatsu:2020sup,Gombor:2020kgu,Gombor:2020auk} and three-point functions involving determinant operators \cite{Jiang:2019xdz,Jiang:2019zig,Yang:2021hrl}, can be mapped to the overlap of a boundary state and an energy eigenstate in integrable systems. Generalizing the techniques developed in the past, exact computations of these observables have been achieved recently \cite{Jiang:2019xdz,Jiang:2019zig,Komatsu:2020sup,Gombor:2020kgu,Gombor:2020auk}.

In all these fascinating developments, there is one corner that has been left out and somehow received little attention. In 2d CFT, there is another class of states called {\it crosscap states} which are just as important as the boundary states. Geometrically they correspond to non-orientable surfaces such as $\mathbb{RP}^2$ and the Klein bottle and have been analyzed extensively in the studies of 2d CFTs on these surfaces \cite{Ishibashi:1988kg,Sagnotti:1987tw,Fioravanti:1993hf,Pradisi:1996yd,Angelantonj:2002ct}. Surprisingly, however, analogous studies for integrable field theories and spin chains have been lacking. The aim of this paper is to fill this gap and initiate a systematic investigation of crosscap states in integrable field theories and spin chains.

The main takeaways of our work are
\begin{itemize}
\item Integrability survives in the presence of crosscaps: The overlaps with the crosscap state, to be called the {\it $p$-functions}, can be computed analytically.
\item The notion of integrability is defined often in the infinite volume limit, be it factorized $S$-matrices, reflection matrices, or form factors. In contrast, the crosscap state is a genuinely finite-volume quantity since it identifies fields at antipodal points and does not admit the infinite volume limit.  It is probably the first example of such a quantity to which integrability is applicable.
\item For general integrable quantum field theories with $\mathbb{Z}_2$-symmetry, one can study their $\mathbb{Z}_2$-orbifolds and fermionization by modifying their (thermodynamic) Bethe ansatz by signs. The $p$-function is sensitive to such modifications.
\end{itemize}
The summary of the rest of the paper is the following:
\begin{itemize}
\item[] {\bf Section \ref{sec:field}}: We compute the overlap between the crosscap state and an arbitrary excited state in integrable field theories with a diagonal $S$-matrix. For the ground state, this gives a crosscap analog of the boundary entropy \cite{Affleck:1991tk} (or equivalently the $g$-function), and we therefore call it the {\it crosscap entropy}. See \eqref{eq:crosscapentropy} for the definition.
We compute it by analyzing the partition function on the Klein bottle and expanding it in two different channels. The result is a ratio of the Fredholm determinants and agrees with the so-called ``universal part''\footnote{This is more like a technical term used in the integrability literature and should not be confused with the standard use of the word ``universal'' in quantum field theory, which refers to the scheme independence or the absence of the counter-term ambiguity. In fact, the boundary entropy, which is independent of the counter term ambiguity, receives contributions both from the ``universal part'' and the ``non-universal prefactor''.} of the $g$-function in integrable field theories up to simple factors. The main difference from the $g$-function is that it only depends on the bulk data and the ``non-universal prefactor'' is absent.
\item[] {\bf Section \ref{sec:flow}}: We compute the crosscap entropy for the integrable RG flow of the staircase model, which contains flows between diagonal unitary minimal models ($A$-series) as its limit. We next show that the flows between the $D$-series minimal models can be obtained by the $\mathbb{Z}_2$-orbifold of the staircase model. The result for the $A$-series decreases monotonically along the RG flow while that of the $D$-series starts to increase in the deep infrared, in the vicinity of the $\mathbb{Z}_2$-symmetry broken phase. Finally we explain how to perform fermionization of integrable field theories at the level of the Bethe ansatz. 
\item[] {\bf Section \ref{sec:spin}}: We define an analog of the crosscap states in the XXX spin chain and the non-compact $SL(2,R)$ spin chain. We show that the overlaps with the Bethe eigenstates are given simply by a ratio of determinants, without extra prefactors. These states are long-range entangled and potentially provide interesting initial conditions for quantum quench, which are quite different from short-ranged entangled initial conditions given by the boundary states. We also prove that the crosscap state in the XXX spin chain is annihilated by infinitely many conserved charges.
\end{itemize}
We then discuss future directions in section \ref{sec:conclusion}.
\section{Exact $\boldsymbol{p}$-Function\label{sec:field}}
In this section, we generalize the derivation of exact $g$-functions in integrable field theories to overlaps between the crosscap state and arbitrary excited states. In subsection \ref{subsec:klein}, we discuss general properties of the crosscap state and the partition function on the Klein bottle. We also give a definition of the crosscap entropy and explain its relation to the crosscap overlap. In subsection \ref{subsec:computation}, we compute the crosscap overlap in integrable field theories. Throughout this section, we assume that the theory is parity-invariant and excitations are scalar.

\begin{figure}[h]
\centering
\begin{minipage}{0.49\hsize}
\centering
\includegraphics[clip, height=3cm]{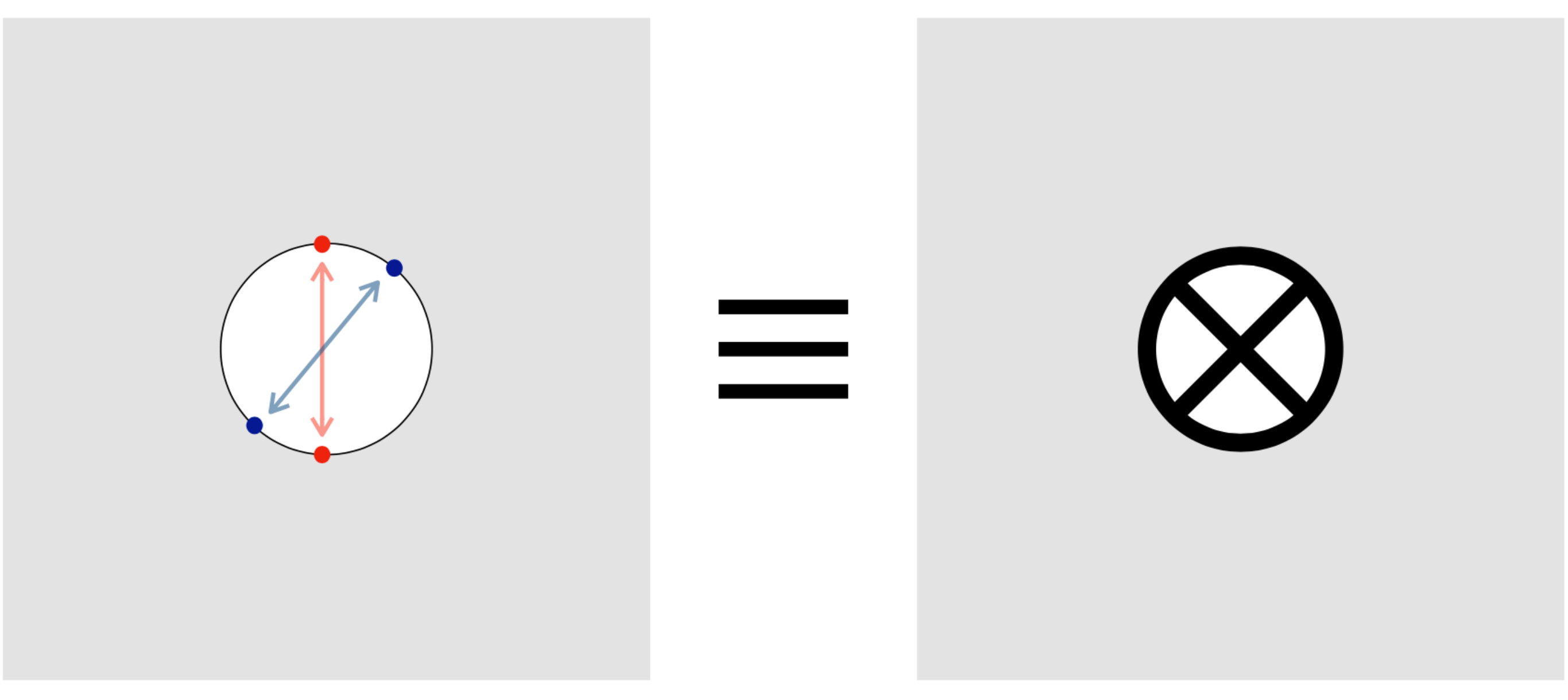}\\
(a)
\end{minipage}
\begin{minipage}{0.49\hsize}
\centering
\includegraphics[clip, height=3cm]{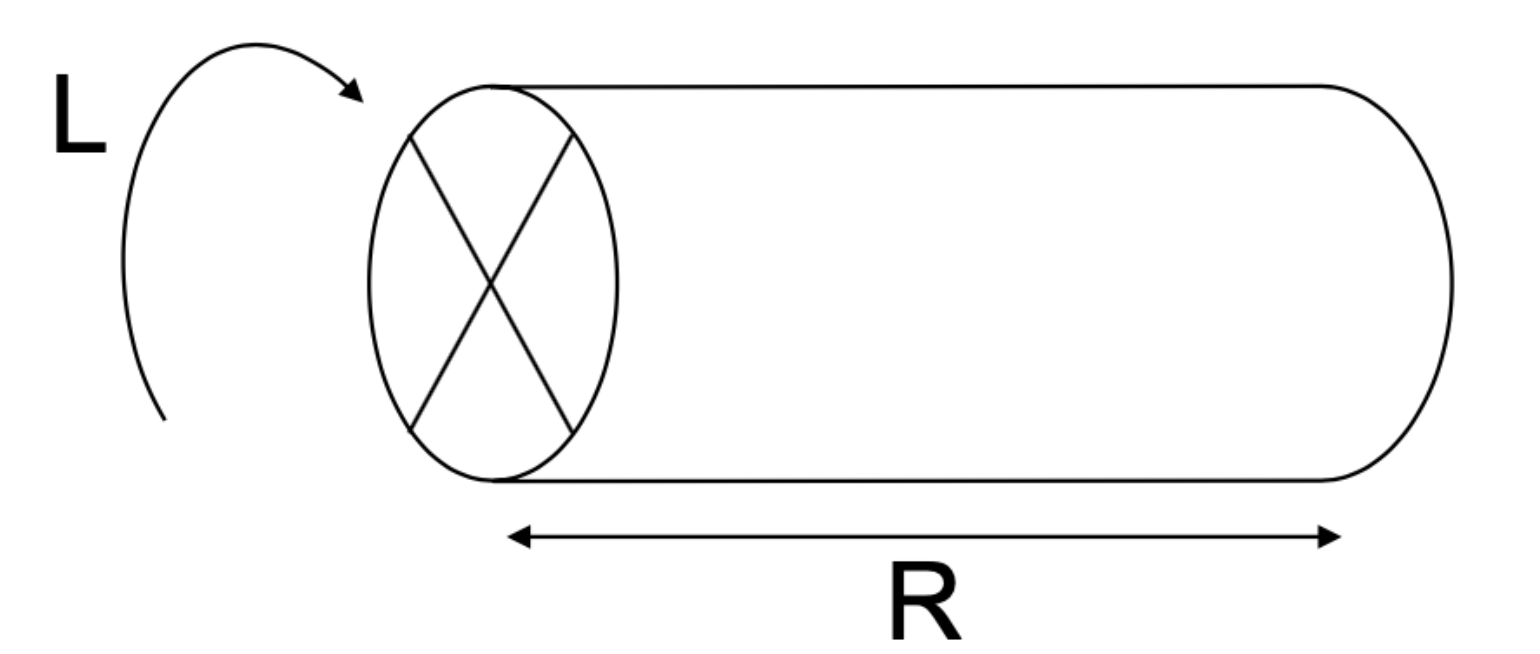}\\
(b)
\end{minipage}
\caption{Definitions of the crosscap (a) and the Klein bottle (b).}\label{fig:defcross}
\end{figure}\vspace{-15pt}
\subsection{Klein bottle and crosscap entropy\label{subsec:klein}}
To define crosscaps, we cut out a disk from a two-dimensional surface and identify antipodal points on the boundary of the disk (see figure \ref{fig:defcross}-(a)). This manipulation makes the surface non-orientable and the state created by this procedure is called the {\it crosscap state}. Two commonly-studied closed non-orientable surfaces are $\mathbb{RP}^2$ and the Klein bottle. They can be obtained by inserting one or two crosscap states on $S^2$ respectively. The crosscap states were studied extensively in 2d CFT, where part of the motivation came from the analysis of string theory in orientifold spacetimes \cite{Sagnotti:1987tw,Fioravanti:1993hf,Pradisi:1996yd,Angelantonj:2002ct}.

To compute the crosscap overlaps, we consider a cylinder of length $R$ and circumference $L$ and contract the two ends with the crosscap states (see figure \ref{fig:defcross}-(b)). This makes the surface topologically equivalent to the Klein bottle. As mentioned above, the Klein bottle can also be obtained by inserting two crosscaps on $S^2$, but here it is important to start with the cylinder, which is locally flat, since our interest is in massive QFT, not CFT. 

The partition function of this Klein bottle $Z_{\mathbb{K}}(R,L)$ can be expanded in two different channels, depending on either we view $R$ or $L$ as the (imaginary) time direction. If we take $R$ as the time direction, we obtain an expansion
\beq\label{eq:crosscapchannel}
\begin{aligned}
Z_{\mathbb{K}}(R,L)&=\sum_{\psi_L}e^{-E_{\psi_L}R}\left|\langle \mathcal{C}|\psi_{L}\rangle\right|^2\overset{R\to\infty}{=}e^{-E_{\Omega_{L}}R}\left|\langle \mathcal{C}|\Omega_{L}\rangle\right|^2+\cdots\period
\end{aligned}
\eeq
Here $\psi_\ell$ is the state defined on the spatial length $\ell$, $|\mathcal{C}\rangle$ is the crosscap state, and $\Omega$ is the ground state. In the literature, this channel is often called the {\it tree channel}.

\begin{figure}[t]
\centering
\includegraphics[clip, height=3.5cm]{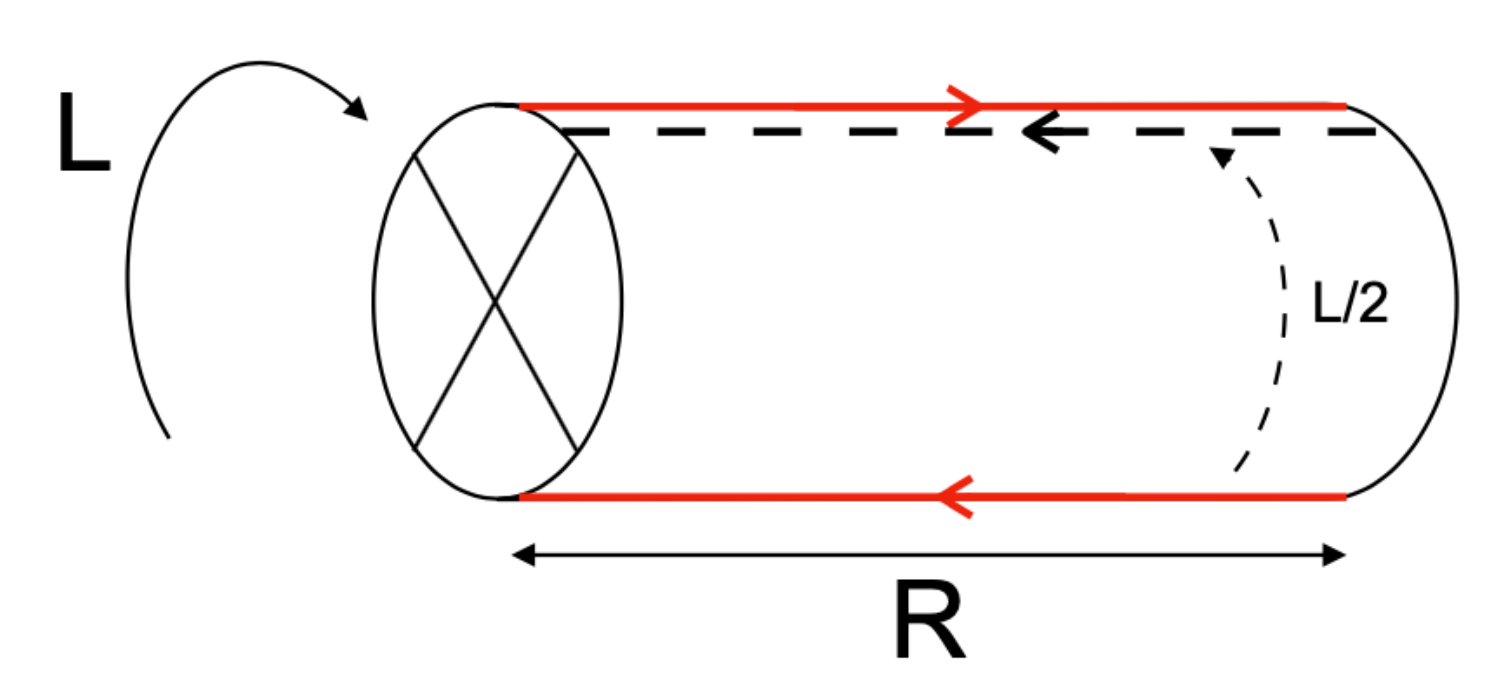}
\caption{Expansion of the Klein bottle partition function in the loop channel. Because of the antipodal identification, the Hilbert space is defined on a union of two red lines, which together form a circle of length $2R$. 
After the time evolution of $L/2$, a state represented by the dashed line gets identified with its parity image represented by the top red line.}
\label{fig:loopchannel}
\end{figure}
The expansion in the other channel (called the {\it loop channel}) is slightly more complicated (see figure \ref{fig:loopchannel}). Owing to the antipodal identification at the boundary of the cylinder, the Hilbert space in the other channel is defined on a circle of length $2R$ not $R$. As can be seen in the figure, the states defined on this circle get identified with their parity images after the time evolution for a period $L/2$. This leads to an expression
\beq\label{eq:parityweightedtrace}
Z_{\mathbb{K}}(R,L)={\rm Tr}_{2R}\left[\Pi \, e^{-HL/2} \right]=\sum_{\psi_{2R}}e^{-E_{\psi_{2R}}\,L/2}\,\,\langle\psi_{2R}|\Pi|\psi_{2R} \rangle\period
\eeq
Here $\Pi$ is the parity operator while $H$ is the Hamiltonian. Re-organizing the sum in terms of eigenstates of the parity, we can rewrite it as
\beq\label{eq:signedpartition}
Z_{\mathbb{K}}(R,L)=\sum_{\psi_{2R}}\epsilon_{\psi_{2R}}e^{-E_{\psi_{2R}}\,L/2}\comma
\eeq
where $\epsilon_{\psi}$ is the eigenvalue of the parity for the state $\psi$, which takes either $+1$ or $-1$. The equality of the two expressions \eqref{eq:crosscapchannel} and \eqref{eq:signedpartition} in the large $R$ limit gives
\beq\label{eq:channelduality}
\lim_{R\to\infty}Z_{\mathbb{K}}(R,L)=\lim_{R\to\infty}\left[\sum_{\psi_{2R}}\epsilon_{\psi_{2R}}e^{-E_{\psi_{2R}}\,L/2}\right]\simeq  e^{-E_{\Omega_{L}}R}\left|\langle \mathcal{C}|\Omega_{L}\rangle\right|^2\period
\eeq
This shows that the overlap $\langle \mathcal{C}|\Omega_{L}\rangle$ controls the density of states weighted by the parity $\epsilon_{\psi}$.  To make this statement more precise, we consider the {\it parity-weighted free energy}
\beq
F_{\mathbb{K}}\equiv-\lim_{R\to\infty}\log Z_{\mathbb{K}}(R,L)\period
\eeq
Without the parity weight $\epsilon$, this would give a definition of a thermal free energy in the infinite volume limit $(R\to \infty)$. Now, using the relation \eqref{eq:channelduality}, we find that $F_{\mathbb{K}}$ behaves as
 \beq
 F_{\mathbb{K}}=RE_{\Omega_L}-\log\left[|\langle \mathcal{C}|\Omega_{L}\rangle|^2\right]+O(1/R)\period
 \eeq 
 This shows that the parity-weighted free energy contains an $O(1)$ term in addition to the usual extensive contribution proportional to the volume $2R$. The structure is reminiscent of the thermal free energy of a system with boundaries, for which the {\it boundary entropy}, also known as the $g$-function, gives an $O(1)$ contribution. The boundary entropy is defined in terms of the overlap with the boundary state $|\mathcal{B}\rangle$ as
 $s_{\mathcal{B}}=(1-L\del_L)\log |\langle \mathcal{B}|\Omega_L\rangle|$.
 Based on the similarity, we call the following quantity the {\it crosscap entropy}:
 \beq\label{eq:crosscapentropy}
 s_{\mathcal{C}}=\log| p|  \qquad \qquad p\equiv \langle \mathcal{C}|\Omega_{L}\rangle\period
 \eeq
The crosscap entropy for rational conformal field theories was discussed already by Tu in \cite{Tu:2017wks} (see also \cite{Garcia-Compean:2018ury}) and it was later extended to compactified boson CFTs \cite{Tang:2018mgw}. The main goal of this paper is to discuss it away from fixed points in particular in integrable QFTs.

Let us make a few remarks on the definition of the crosscap entropy \eqref{eq:crosscapentropy}. 
\begin{itemize}
\item For the boundary entropy, the subtraction of the $L\del_{L}$ term removes non-universal contributions related to counter terms localized at the boundary, and is crucial for defining a quantity that is scheme-independent and monotonically decreases along the RG flow \cite{Cuomo:2021rkm}. For the crosscaps, we do not expect such counter terms\footnote{The counter terms that one can write for the Klein bottle partition function give extensive contributions; namely they are proportional to $RL$, which can be removed by taking a ratio with a torus partition function as discussed in \cite{Tu:2017wks}.} and therefore we defined the entropy without subtracting the $L\del_{L}$ term. We will study the monotonicity of $s_{\mathcal{C}}$ in examples in the next section. 
\item Here we focused on the crosscap overlap for the ground state. As we see below, the crosscap overlaps can also be computed for any excited state in integrable field theories. They do not have a natural thermodynamic interpretation, but are nevertheless important observables in integrable theories on non-orientable surfaces.
\item In string theory and 2d CFT \cite{Ishibashi:1988kg,Sagnotti:1987tw,Fioravanti:1993hf,Pradisi:1996yd,Angelantonj:2002ct}, one often considers more general crosscap states for which the parity $\Pi$ in \eqref{eq:parityweightedtrace} is replaced with a product of $\Pi$ and some other $\mathbb{Z}_2$ transformation. We will not discuss such a generalization in this paper, but it is an interesting direction to explore. We will come back to this point in the conclusion.
\end{itemize}
\subsection{Crosscap overlaps in integrable field theories\label{subsec:computation}}
We now compute the crosscap overlaps in integrable field theories. To be concrete, we consider theories with a single species of particles without bound states, the prototypical example being the sinh-Gordon model.

Before proceeding to the calculation, let us briefly summarize the state of research of the $g$-function, since the discussion below heavily  relies on it. The first complete proposal for the $g$-function in integrable field theories with diagonal scattering was made in \cite{Dorey:2004xk}, building on earlier works \cite{LeClair:1995uf,Woynarovich:2004gc}. The proposal was verified later in \cite{Pozsgay:2010tv}, which provided a streamlined derivation based on the thermodynamic Bethe ansatz \cite{Zamolodchikov:1989cf}. These results were recently generalized in \cite{Kostov:2018dmi} to non-diagonal scattering, and a further reformulation was made in \cite{Kostov:2019sgu} which revealed a underlying effective field theory description whose path integral localizes to the saddle point. More recently, the generalization of the $g$-function to excited states was achieved in \cite{Jiang:2019xdz,Jiang:2019zig}, based on the analytic continuation approach developed in \cite{Dorey:1996re} for the spectrum. In what follows, we generalize these techniques to the crosscap overlaps.

\paragraph{Derivation.}The starting point of our analysis is the relation \eqref{eq:channelduality}, which we display here again in a slightly different form:
\beq\label{eq:keyformula}
\lim_{R\to\infty}{\rm Tr}_{2R}\left[\Pi \, e^{-\hat{H}L/2}\right]\simeq e^{-E_{\Omega} R}\left|\langle \mathcal{C}|\Omega_{L}\rangle\right|^2\period
\eeq
In integrable theories, the energy eigenstates in the infinite volume limit ($R\to\infty$) can be described as a collection of excitations, and are labelled by a set of momenta $|\{p_j\}\rangle$ ($j=1,\ldots, M$) that satisfy the Bethe equations
\beq\label{eq:bethe}
1=e^{2ip_jR}\prod_{k\neq j}S(p_j,p_k)\period
\eeq
The parity transformation simply flips the signs of these momenta,
\beq\label{eq:propto}
\Pi|\{p_j\}\rangle\propto|\{-p_j\}\rangle\period
\eeq
Note that the constant of proportionality depends on the definition of the state $|\{p_j\}\rangle$ and cannot be determined just from this argument. (We will see shortly that the constant of proportionality is $1$ in the standard normalization of the Bethe wave function.)
Using \eqref{eq:propto}, we conclude that the states whose momenta are {\it not} invariant under the sign flip\footnote{In the literature, the  states which are invariant under the sign flip of momenta are often called {\it unpaired states} while the states which are not invariant are called the {\it paired states} (simply because such states come in pairs). The unpaired states are known to be annihilated by infinitely many odd spin charges. See \cite{Kristjansen:2010kg} for discussions in the context of spin chain.} do not contribute in the parity-weighted trace:
\beq
\langle \{p_j\}|\Pi|\{p_j\}\rangle\propto\langle \{p_j\}|\{-p_j\}\rangle=0\qquad \text{if $\{p_j\}\neq \{-p_j\}$}\period 
\eeq
In other words, these states come in pairs $|\{p_j\}\rangle\pm \Pi|\{p_j\}\rangle$, and their contributions cancel out in the weighted trace.

The next step is to show that all the states whose momenta {\it are} invariant under the sign flip have eigenvalue $+1$ under the parity. To see this, consider a coordinate wave function
\beq\label{eq:bethewave}
|\{p_1,\ldots, p_M\}\rangle=\int_{x_1<\cdots < x_M} dx_1\cdots dx_M \Psi_{p_1,\ldots, p_M}(x_1,\ldots, x_M) |x_1,\ldots, x_M\rangle\period
\eeq
Here $|x_1,\ldots, x_M\rangle$ denotes a state in which the $j$-th excitation is at position $x_j$.
When the excitations are far apart, the wave function is given by a sum over plane waves 
\beq\label{eq:waveasympt}
 \Psi_{p_1,\ldots, p_M}(x_1,\ldots, x_M) \overset{x_1\ll \cdots \ll x_M}=e^{i(p_1 x_1+p_2x_2+\cdots )}+S(p_1,p_2)e^{i(p_2 x_1+p_1x_2+\cdots )}+\cdots \period
\eeq
Acting the parity transformation to \eqref{eq:bethewave}, we obtain
\beq
\Pi|\{p_1,\ldots, p_M\}\rangle=\int_{x_1<\cdots < x_M} dx_1\cdots dx_M \Psi_{p_1,\ldots, p_M}(x_1,\ldots, x_M) |2R-x_M,\ldots, 2R-x_1\rangle\period
\eeq
Changing the integration variables from $x_j$ to $x_j^{\prime}=2R-x_{M+1-j}$, we can rewrite this as
\beq
\Pi|\{p_1,\ldots, p_M\}\rangle=\int_{x_1^{\prime}<\cdots < x_M^{\prime}} dx_1^{\prime}\cdots dx_M^{\prime} \Psi_{p_1,\ldots, p_M}(2R-x_M^{\prime},\ldots, 2R-x_1^{\prime}) |x_1^{\prime},\ldots, x_M^{\prime}\rangle\period
\eeq
Now using the asymptotic form of the wave function \eqref{eq:waveasympt} and the parity invariance of the $S$-matrix $S(p,q)=S(-q,-p)$, we can check that
\beq
\Psi_{p_1,\ldots, p_M}(2R-x_M^{\prime},\ldots, 2R-x_1^{\prime})=e^{2i R\sum_{k}p_k}\Psi_{-p_M,\cdots, -p_1}(x_1^{\prime},\cdots, x_M^{\prime})\period
\eeq
Since the Bethe equation \eqref{eq:bethe} implies $e^{2iR\sum_{k}p_k}=1$, we conclude that the state $|\{p_j\}\rangle$ transforms under the parity as
\beq
\Pi|\{p_1,\ldots, p_M\}\rangle=|\{-p_M,\ldots, -p_1\}\rangle\period
\eeq
In particular, this means that the state with a parity-invariant set of momenta 
\beq
\{p_1,\ldots, p_M, -p_M, \ldots, -p_1\} \qquad \text{or}\qquad \{p_1,\ldots, p_M, 0, -p_M, \ldots, -p_1\}
\eeq
has an eigenvalue $+1$ under the parity.

Therefore the left hand side of \eqref{eq:keyformula} reduces to a restricted thermal sum over states
\beq\label{eq:reducedthermal}
{\rm Tr}_{2R}\left[\Pi \, e^{-\hat{H}L/2}\right]=\sum_{\{p_j\}=\{-p_j\}}e^{-\frac{L}{2}\sum_{j}E(p_j)}\comma
\eeq
where the sum $\sum_{\{p_j\}=\{-p_j\}}$ is taken over solutions to the Bethe equation \eqref{eq:bethe} that are invariant under the sign flip. In the large volume limit $R\to\infty$, this sum can be evaluated using the standard trick of the thermodynamic Bethe ansatz (TBA)---namely we replace the sum over states with a path integral of density of excitations and evaluate it around the saddle point\footnote{The result is given by a product of an exponentially decaying piece which comes from the evaluation of the action at the saddle point, and the $O(1)$ piece that comes from the one-loop fluctuations around the saddle point. The former gives the ground energy in the tree channel while the latter gives the crosscap overlap. For details, see e.g.~\cite{Jiang:2019xdz}.}. The only modification from the standard TBA is the constraint $\{p_j\}=\{-p_j\}$.

To deal with the constraint, let us study the Bethe equation for a parity-invariant set of momenta. It takes a different form depending on whether we have a zero-momentum particle, and for a reason that will become clear later, we call the two sectors ${\bf S}$ and ${\bf T}$:
\begin{align}
&{\bf S}:\qquad 1=e^{2ip_jR}S(p_j,-p_j)\prod_{k\neq j}S(p_j,p_k)S(p_j,-p_k)\comma\label{eq:Sparbethe}\\
&{\bf T}:\qquad 1=e^{2i p_j R}S(p_j,-p_j)S(p_j,0)\prod_{k\neq j}S(p_j,p_k)S(p_j,-p_k)\period\label{eq:Tparbethe}
\end{align}
Now comes the crucial observation. These equations take the same form as the Bethe equation for a system with two identical boundaries\footnote{For two non-identical boundaries, the total reflection phase is given by a product of the left and the right reflection phases $R_{L}(-p)R_{R}(p)$. Here we used the relation $R_{L}(-p)=R_{R}(p)$, which holds for identical boundaries.}
\beq\label{eq:boundaryBethe}
1=e^{2ip_jR}\left(R(p_j)\right)^2\prod_{k\neq j}S(p_j,p_k)S(p_j,-p_k)\comma
\eeq
if we identify the reflection matrix $R(p)$ and $S$-matrices as follows: 
\beq\label{eq:repl}
\left(R(p_j)\right)^2\leftrightarrow \begin{cases}S(p_j,-p_j)\qquad& :{\bf S}\\S(p_j,-p_j)S(p_j,0)\qquad &:{\bf T}\end{cases} \period
\eeq 
Moreover, the parity-weighted sum \eqref{eq:reducedthermal} can be rewritten in the following way,
\beq\label{eq:thermalST}
{\rm Tr}_{2R}\left[\Pi \, e^{-\hat{H}L/2}\right]=\sum_{{\bf S}}e^{-L\sum_{p_j>0}E(p_j)}+e^{-\frac{mL}{2}}\sum_{{\bf T}}e^{-L\sum_{p_j>0}E(p_j)}\comma
\eeq
with $m\equiv E(p=0)$ being the mass of the particle.
This shows that the contribution from each sector takes the same form as the thermal partition function of the boundary problem with the inverse temperature $L$ (up to a prefactor $e^{-mL/2}$ in the ${\bf T}$-sector). 
 These imply that we can recycle the results for the boundary problem, in particular the result for the $g$-function. 

\paragraph{Result.}In order to write down the result for the crosscap overlap, let us recall the result for the $g$-function \cite{Pozsgay:2010tv,Kostov:2018dmi,Jiang:2019xdz} (see e.g. section 6.2.2 of \cite{Jiang:2019xdz} for the derivation),
\beq\label{eq:g-func}
\begin{aligned}
|\langle \mathcal{B}|\Omega_L\rangle|^2=\exp\left[2\int_{0}^{\infty}\frac{du}{2\pi}\Theta (u)\log (1+Y(u))\right]\frac{\det \left[1-\hat{G}_{-}\right]}{\det \left[1-\hat{G}_{+}\right]}\period
\end{aligned}
\eeq
Here $Y(u)$ is the $Y$-function\footnote{In the literature, one often writes the results in terms of the ``pseudo energy'' $\epsilon(u)$: $Y(u)=e^{-\epsilon (u)}$.} and it satisfies the TBA equation
\beq\label{eq:TBAequation}
0=LE(u)+\log Y (u)-\log (1+Y)\ast\mathcal{K}_{+}(u)\comma
\eeq
with $E(u)$ being the energy. The notation $A\ast B$ stands for the convolution $\int_0^{\infty}\frac{dv}{2\pi}A(u,v)B(v)$ while the kernels $\mathcal{K}_{\pm}$ are defined in terms of the $S$-matrix $S(u,v)$ as
\beq
\mathcal{K}_{\pm}(u,v)=\frac{1}{i}\del_u\left[\log S(u,v)\pm \log S(u,-v)\right]\period
\eeq
The Fredholm determinants $\det \left[1-\hat{G}_{\mp}\right]$ are defined in terms of the kernels $\mathcal{K}_{\pm}$ as
\beq
\hat{G}_{\pm}\cdot f(u)=\int^{\infty}_{0}\frac{dv}{2\pi}\frac{\mathcal{K}_{\pm}(u,v)}{1+1/Y(v)}f(v)\period
\eeq
The only factor which depends on the reflection matrix $R(u)$ is $\Theta (u)$ and it reads
\beq\label{eq:eqforTheta}
\Theta (u)=\frac{1}{i}\del_u\log R(u)-\pi \delta (u)-\frac{1}{i}\del_u\left.\log S(u,v)\right|_{v=-u}\period
\eeq
Here, the subtraction of the delta function $-\pi\delta (u)$ is necessary because the boundary Bethe equation \eqref{eq:boundaryBethe} allows $p_j=0$ as a formal solution although the state with $p_j=0$ does not exist as a physical state.

Let us now apply the formula to the crosscap state. For the ${\bf S}$-sector, we first perform the replacement \eqref{eq:repl} and use the identity
\beq
\del_u\log R(u)\qquad \mapsto \qquad \frac{1}{2}\del_u\log S(u,-u)=\left.\del_u \log S(u,v)\right|_{v=-u}\period
\eeq
A small difference from the boundary problem is that the Bethe equation \eqref{eq:Sparbethe} does not admit $p_j=0$ as a solution since the right hand side of \eqref{eq:Sparbethe} evaluates to $-1$, not $1$, upon setting $p_j=0$ thanks to $S(0,0)=-1$. Thus, the subtraction $-\pi\delta(u)$ is not necessary. As a result, the prefactor $\Theta$ vanishes and the contribution from the {\bf S}-sector is given simply by
\beq\label{eq:Svac}
\begin{aligned}
|\langle \mathcal{C}|\Omega_L\rangle|^2\quad \overset{{\bf S}}{\supset}\quad \frac{\det \left[1-\hat{G}_{-}\right]}{\det \left[1-\hat{G}_{+}\right]}\period
\end{aligned}
\eeq

On the other hand, the contribution from the {\bf T}-sector can be evaluated by performing the replacement
\beq
\frac{1}{2}\del_u\log R(u)\quad \mapsto \quad\left.\del_u \log S(u,v)\right|_{v=-u} +\frac{\del_u \log S(u,0)}{2}\period
\eeq
Unlike the {\bf S}-sector, the Bethe equation for the {\bf T}-sector {\it does} admit $p_j=0$ as a solution. This however should not be included in the TBA computation since we already separated out the contribution from the zero-momentum excitation in \eqref{eq:thermalST} and we cannot have two excitations with the same momentum. Therefore we need to subtract $\pi \delta(u)$ in the final answer.
Combined with the extra factor $e^{-mL/2}$ in \eqref{eq:thermalST}, this leads to an expression
\beq\label{eq:Timtermediate}
\begin{aligned}
&|\langle \mathcal{C}|\Omega_L\rangle|^2\quad \overset{{\bf T}}{\supset}\\
& \exp\left[-\frac{mL}{2}+\frac{1}{2}\int_0^{\infty}\frac{du}{2\pi}\mathcal{K}_{+}(0,u)\log (1+Y(u))\right]\frac{\det \left[1-\hat{G}_{-}\right]}{\sqrt{1+Y(0)}\det \left[1-\hat{G}_{+}\right]}\period
\end{aligned}
\eeq
Here the square-root in $1/\sqrt{1+Y(0)}$ comes from $\int_0^{\infty}du\,\delta(u)f(u)=\frac{1}{2}f(0)$, and we used\footnote{The second equality comes from $ \del_u\log S(u,v)-\del_v\log S(v,u)=\del_u \log S(u,v)+\del_u \log S(v,u)=0$.}
\beq
\frac{1}{i}\del_u \log S(u,0)=\frac{1}{2}\mathcal{K}_{+}(u,0)=\frac{1}{2}\mathcal{K}_{+}(0,u)\period
\eeq
The exponent in \eqref{eq:Timtermediate} can be simplified using the TBA equation \eqref{eq:TBAequation} evaluated at $u=0$. The result reads
\beq
|\langle \mathcal{C}|\Omega_L\rangle|^2\quad \overset{{\bf T}}{\supset}\qquad \sqrt{\frac{Y(0)}{1+Y(0)}}\frac{\det \left[1-\hat{G}_{-}\right]}{\det \left[1-\hat{G}_{+}\right]}\period
\eeq

Summing the two contributions and taking the square root, we arrive at our main formula
\beq\label{eq:finalvac}
|p|=\left|\langle \mathcal{C}|\Omega_L\rangle\right|=\sqrt{\left(1+\sqrt{\frac{Y(0)}{1+Y(0)}}\right)\frac{\det \left[1-\hat{G}_{-}\right]}{\det \left[1-\hat{G}_{+}\right]}}\period
\eeq
As compared to the formula for the $g$-function, it does not contain a ``non-universal'' prefactor that depends on the reflection matrices. Thus, at the level of the formula, the crosscap overlap can be viewed as the ``simplest possible $g$-function''. Another comment is that the result \eqref{eq:finalvac} only gives the absolute value of the overlap. In general, we expect the phase of the overlap to carry important physical information\footnote{For instance, for orientifolds in string theory, the overlap can be negative reflecting the negative tension of the orientifold plane.} as well.  However its computation requires more careful analysis and we leave it for future studies.
\paragraph{Excited states and asymptotic limit.}We can also derive the generalization for excited states applying the argument given in \cite{Jiang:2019xdz}, which uses the analytic-continuation trick developed by Dorey and Tateo for the spectrum \cite{Dorey:1996re}. An immediate consequence of this is that the crosscap state has non-zero overlaps only with states that are parity-symmetric; namely states whose rapidities are invariant under the parity transformation. In the case of integrable boundary states, this selection rule reflects the fact that the boundary states preserve infinitely many parity-odd conserved charges \cite{Ghoshal:1993tm,Piroli:2017sei}. We expect the same to hold for crosscap states although we do not present a general proof in this paper. (We do present a proof of an analogous statement for the XXX spin chain in section \ref{sec:spin}.)

The result of the analytic continuation is (see section 6.3 of \cite{Jiang:2019xdz} for the derivation):
\beq\label{eq:fieldfinal}
|\langle \mathcal{C}|\Psi_{L}\rangle|=\sqrt{\left(1+\sqrt{\frac{Y(0)}{1+Y(0)}}\right)\frac{\det \left[1-\hat{G}_{-}^{\bullet}\right]}{\det \left[1-\hat{G}_{+}^{\bullet}\right]}}\period
\eeq
Here $|\Psi_{L}\rangle$ is a parity-symmetric excited state satisfying the excited-state TBA
\beq
\begin{aligned}
&0=\log Y(u)+LE(u)+\sum_{k}\log \left(S(\tilde{u}_k,u)S(\tilde{u}_k,-u)\right)-\log (1+Y)\ast \mathcal{K}_{+}(u)\comma
\end{aligned}
\eeq
with $\tilde{u}_k$'s satisfying the exact Bethe equation $1+Y(\tilde{u}_k)=0$. The deformed Fredholm determinants $\det \left[1-\hat{G}_{\pm}^{\bullet}\right]$ are given by a combination of sums and convolutions:
\beq
\hat{G}^{\bullet}_{\pm}\cdot f(u)=\sum_{k}\frac{i\mathcal{K}_{\pm}(u,u_k)}{\del_u \log Y(\tilde{u}_k)}f(\tilde{u}_k)+\int^{\infty}_{0}\frac{dv}{2\pi}\frac{\mathcal{K}_{\pm}(u,v)}{1+1/Y(v)}f(v)\period
\eeq

With the formula for the excited states, we can analyze the large volume limit $L\to \infty$, which is often called the {\it asymptotic limit}. In the limit, states that have non-zero overlaps are labelled by a parity-symmetric set of rapidities 
\beq
{\bf u}=\{u_1,\ldots, u_{M}\}\qquad\qquad  u_{j+\frac{M}{2}}=-u_{j}\comma
\eeq
satisfying the Bethe equation
\beq
1=e^{ip(u_j)L}\prod_{k\neq j}^{M}S(u_j,u_k)\period
\eeq
In addition, in the formula for the overlap, the $Y$-function is exponentially suppressed on the real axis and the terms that involve the convolution can be dropped. The result, after rewriting, reads (see section 6.3.2 of \cite{Jiang:2019xdz} for details)
\beq\label{eq:asymptotic}
|\langle \mathcal{C}|\Psi_{L}\rangle|\overset{L\to\infty}{=}\sqrt{\frac{\det G_{+}}{\det G_{-}}}\comma
\eeq
with
\beq\label{eq:gpm0}
\left(G_{\pm}\right)_{1\leq i,j\leq\frac{M}{2}}=\left[L\del_{u}p(u_i)+\sum_{k=1}^{\frac{M}{2}}\mathcal{K}_{+}(u_i,u_k)\right]\delta_{ij}-\mathcal{K}_{\pm}(u_i,u_j)\period
\eeq
Later in section \ref{sec:flow}, we will see exactly the same structure as \eqref{eq:asymptotic} in integrable spin chains. 
\section{Crosscap Entropy in Integrable Flow\label{sec:flow}}
In this section, we use the formulae derived in the previous section to study the behaviour of the $p$-function under the RG flow and we specifically apply it to the so-called staircase model first introduced by Al.~Zamolodchikov \cite{Zamolodchikov:2006vf}. 
In subsection \ref{sec:dseries}, we propose a generalization of the staircase model whose RG flow  interpolates between the $D$-series minimal models obtained by $\mathbb{Z}_2$-orbifolding the $A$-series minimal models. 
Finally, in subsection \ref{sec:fermi}, we discuss the fermionization of integrable field theories and how to accomodate them in the Bethe ansatz. In particular, this allows for yet another generalization of the staircase model.

\subsection{RG flow for the $p$-function in the staircase model}\label{sec:staircase}
\paragraph{Model.}The staircase model is defined abstractly by an $S$-matrix. To this date the corresponding Lagrangian realization remains unknown. Its origin stems from the sinh-Gordon model which is a relativistic integrable theory of a single massive scalar field whose Lagrangian is
\beq
\mathcal{L}_{\rm{shG}} = \frac{1}{2} (\partial \Phi)^2 - \frac{m^2}{b^2} \cosh ( b \Phi )\,.
\eeq 
with $m$ and $b$ being the mass of the elementary field $\Phi$ and the coupling constant respectively.
The sinh-Gordon exact $S$-matrix is given in terms of the difference of rapidities of the corresponding excitations being scattered
\beq
S(u-v) = \frac{\sinh (u-v) - i\sin \gamma}{\sinh (u-v) + i \sin \gamma}\,.
\eeq
Importantly, the parameter $\gamma$ is related to the coupling constant of the model by 
\beq
\gamma=\frac{\pi b^2}{8\pi+b^2}\comma
\eeq
and the $S$-matrix enjoys invariance under the transformation $\gamma \rightarrow \pi-\gamma$. This is the famous weak-strong coupling duality of the sinh-Gordon model. At $\gamma= \pi/2$, the theory is self-dual.

 In \cite{Zamolodchikov:2006vf}, Al.~Zamolodchikov proposed to move away from the self-dual point by the following analytic continuation
\beq
\gamma = \frac{\pi}{2} \pm i \theta_0
\eeq
with $\theta_0$ being a real parameter. The resulting $S$-matrix remains real-analytic \cite{Dorey:1996gd} (i.e. $S(u)$ is real for purely imaginary $u$) besides maintaining the remaining  constraints from the original theory, namely unitarity and crossing symmetry. It is therefore considered to be still a physical $S$-matrix describing the scattering of asymptotic states in some putative massive QFT whose underlying microscopic description remains somewhat obscure. We will take this point of view, and use this model as a playground to study $p$-functions.

\begin{figure}[t]
\centering
\includegraphics[clip, height=6.3cm]{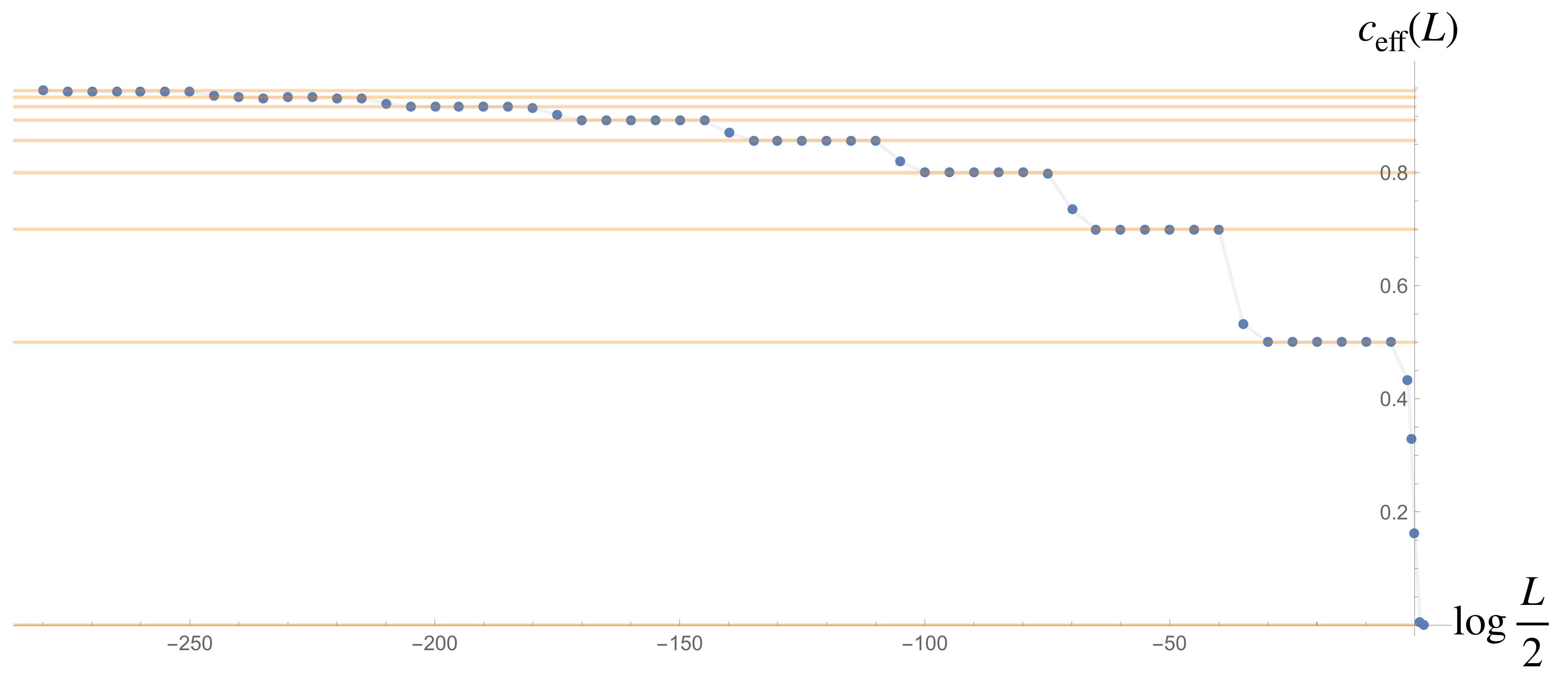}\\
\caption{The dependence of the effective central charge for the staircase model on the volume $L$. The orange horizontal lines indicate the central charges of the minimal models and the effective central charge develops plateaux precisely at those values.}\label{fig:ceff}
\end{figure}

The salient feature of the staircase model occurs when $\theta_0$ is sent to infinity. As $\theta_0$ increases, the effective central charge (or equivalently the ground-state energy) as a function of the volume of the system $L$ develops several \textit{plateaux}, whose approximate locations are at the values of $L$ obeying $\log L \sim -(m-3)\theta_{0}/2$ for integers $m\geq2$, see figure \ref{fig:ceff}. At each of these plateaux, the effective central charge matches precisely the central charge of a diagonal unitary minimal model, also known as the $A$-series minimal models, $\mathcal{M}_{m}^{(A)}$:
\beq
c_{\rm eff}(L)\sim c_m=1-\frac{6}{m(m+1)}\qquad\qquad  \left(\log L \sim -(m-3)\theta_{0}/2\right)\period
\eeq
 The staircase behavior is more pronounced as $\theta_0$ grows and in the limiting behavior as $\theta_0 \rightarrow \infty$, it describes the sequence of transitions $\mathcal{M}^{(A)}_m \rightarrow \mathcal{M}^{(A)}_{m-1}$. These transitions are well-known and amount to the RG flows induced by the least relevant operator $\phi_{13}$ which is part of the spectrum of all these minimal models. This behavior suggests an intepretation for  the staircase model: it describes a parametric family of integrable field theories whose behavior along the RG comes close to that of the unitary minimal models before it flows to a massive theory in the deep infrared.

This is then an example where we can follow the RG behavior using integrability and enjoy  an infinite sequence of benchmark points provided by the unitary minimal models for which we can analytically compute a large amount of information purely from CFT.
\paragraph{Result.}
\begin{figure}[t]
\centering
\includegraphics[clip, height=6.3cm]{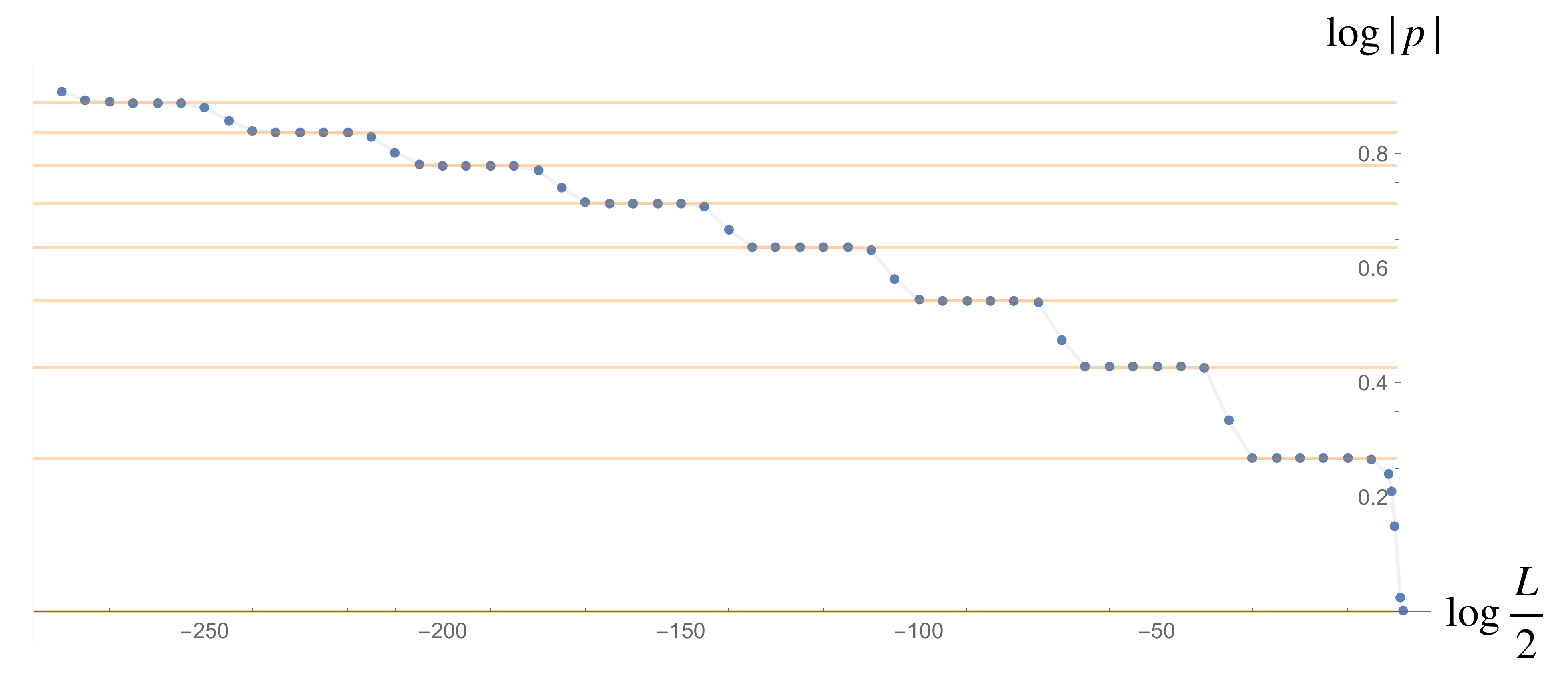}\\
\caption{The $p$-function for the staircase model as a function the volume $L$. Similarly to the effective central charge, it develops plateaux (orange horizontal lines) at the values of the $p$-function for the minimal models in the $A$-series.}\label{fig:logp}
\end{figure}

The result of the numerical evaluation of (\ref{eq:finalvac}) is given in the plot of figure \ref{fig:logp}. The crosscap entropy ($\log |p|$) monotonically decreases with the RG flow and decays to zero in the deep infrared. Along the way, it develops plateaux whose values correspond to the $p$-functions of the minimal models in the $A$-series. 
These plateaux values can be calculated analytically. Referring to the formula (\ref{eq:finalvac}), the (ratio of) Fredholm determinants and the value of $Y(0)$ at the plateau corresponding to $\mathcal{M}_{m}^{(A)}$ were computed in \cite{Dorey:2010ub}. The results read (see \cite{Dorey:2010ub} for derivations)
\begin{align}
\sqrt{\frac{\det\left[1-\hat{G}_{-}\right]}{\det\left[1-\hat{G}_{+}\right]}}&=\begin{cases}\displaystyle{\left(\frac{8}{m(m+1)}\right)^{\frac{1}{4}}\frac{\sin\frac{(m-1)\pi}{2m}}{\sqrt{\sin\frac{\pi}{m}\sin\frac{\pi}{m+1}}}\qquad m:\text{ odd}}\\
\displaystyle{\left(\frac{8}{m(m+1)}\right)^{\frac{1}{4}}\frac{\sin\frac{m\pi}{2(m+1)}}{\sqrt{\sin\frac{\pi}{m}\sin\frac{\pi}{m+1}}}\qquad m:\text{ even}}
\end{cases}\comma\label{eq:plateauratio}\\
Y(0)&=\begin{cases}\left(\cot\frac{\pi}{m+1}\right)^2\qquad &m:\text{ odd}\\
\left(\cot\frac{\pi}{m}\right)^2\qquad &m:\text{ even}
\end{cases}\period\label{eq:plateauy0}
\end{align}
Plugging these values into \eqref{eq:finalvac}, we obtain a closed-form expression for the $p$-function of the $A$-series minimal models:
\beq\label{eq:closedform}
\mathcal{M}^{(A)}_m:\qquad |p|=|\langle \mathcal{C}|\Omega\rangle|=\left(\frac{2}{m(m+1)}\right)^{\frac{1}{4}}\sqrt{\cot\frac{\pi}{2m}\cot\frac{\pi}{2(m+1)}}\period
\eeq
The explicit results for the first few $m$'s are
\beq
\begin{aligned}
&\mathcal{M}^{(A)}_3 :\sqrt{1+\frac{1}{\sqrt{2}}}\comma\qquad&&\mathcal{M}^{(A)}_{4}:\sqrt{1+\frac{1}{\sqrt{2}}}\left(1+\frac{2}{\sqrt{5}}\right)^{\frac{1}{4}}\comma\\
&\mathcal{M}^{(A)}_{5}:\sqrt{2+\sqrt{3}}\left(\frac{5+2\sqrt{5}}{15}\right)^{\frac{1}{4}}\comma
\end{aligned}
\eeq
which correspond to the Ising ($m=3$), the tricritical Ising ($m=4$) and the tetracritical Ising ($m=5$) models\footnote{The results for $m=3,4,5$  appeared before in the literature. See e.g.~\cite{Fioravanti:1993hf,Tsiares:2020ewp,Garcia-Compean:2018ury}.}. To our knowledge, the closed-form expression \eqref{eq:closedform} never appeared before in the literature.
\paragraph{Comparison with CFT.} Let us now check that the values at the plateaux \eqref{eq:closedform}, obtained from integrability, agree with those in CFT. To compute the $p$-function from CFT, we follow the same strategy as in section \ref{sec:field}; namely we consider the Klein bottle partition function, compute it by a parity-weighted sum over states in the loop channel and expand the answer in the tree channel. The additional advantages in working directly with CFTs are
\begin{enumerate}
\item The parity operator exchanges the chiral and the anti-chiral sectors. As a result, only states that survive in the parity-weighted sum are the ones with identical chiral and anti-chiral representations, i.e.~Verma modules corresponding to scalar primaries.
\item The sum over states can be organized into a sum over Virasoro characters. Because of the parity operator which exchanges the chiral and the anti-chiral sectors, it is given by a sum over single Virasoro characters, unlike the torus partition function which is given by a sum of products of two characters (chiral times anti-chiral). 
\item One can rewrite each Virasoro character in the loop channel into a sum of characters in the tree channel using the modular $S$-matrix.
\end{enumerate}
The details of the derivation can be found in the literature \cite{Ishibashi:1988kg,Blumenhagen:2009zz,Tu:2017wks,Tsiares:2020ewp} and we will not repeat it here. The outcome of the analysis is given by the following sum of modular $S$-matrices
\beq\label{eq:generalformulaCFT}
|\langle C|\Omega\rangle|=\left(\sum_{a}n_{a,a}S_{a, I}\right)^{\frac{1}{2}}
\eeq
where $a$ denotes a irreducible representation of the Virasoro algebra, $I$ is the identity representation, $n_{a,b}$ is a degeneracy of states with the representation $a$ for the chiral part and $b$ for the anti-chiral part. $S_{a,b}$ is the modular $S$-matrix between the representations $a$ and $b$.

Before we move on, let us make one clarifying remark on the formula\footnote{We thank Yifan Wang for asking a clarifying question.} \eqref{eq:generalformulaCFT}. When deriving \eqref{eq:generalformulaCFT}, we made an assumption that all the scalar primaries have eigenvalues $+1$ under the parity. This is a reasonable assumption and is in line with what we saw in section \ref{subsec:computation} for integrable field theories. However, strictly speaking one needs to check if this choice is consistent with other conformal bootstrap conditions such as the one coming from the Mobius strip. For the $A$-series minimal models analyzed here, it was shown in \cite{OnogiIshibashi} that this choice is indeed consistent, but an analogous statement is not yet established for the $D$-series minimal models, which we discuss in the next subsection. Therefore, to be precise, we should take the CFT result for the $D$-series minimal model as a conjecture.

To evaluate \eqref{eq:generalformulaCFT} for a given theory, we need to know two data; the modular $S$-matrix and the degeneracy $n_{a,b}$. As is well-known \cite{DiFrancesco:1997nk}, the minimal models are labelled by two coprime integers $(p,q)$ and unitary theories correspond to $(p,q)=(m,m+1)$ or $(m+1,m)$. To list states in these unitary theories in a compact way, we follow \cite{Hsieh:2020uwb} and set $(p,q)=(m,m+1)$ when $m$ is odd and $(p,q)=(m+1,m)$ when $m$ is even. In this convention, the full set of (non-chiral) characters for the $A$-series minimal model reads
\beq\label{eq:repA}
\mathcal{M}_m^{(A)}: \sum_{1\leq r\leq q-1}\sum_{1\leq s\leq \frac{p-1}{2}}\chi_{r,s}\overline{\chi_{r,s}}\period
\eeq
Here $\chi_{r,s}$ is a (chiral) character of a irreducible representation of the Virasoro algebra corresponding to a primary with dimension $L_0=\frac{(pr-qs)^2-1}{4 p q}$.
For later convenience, let us also write down characters for the $D$-series in this convention:
\beq\label{eq:repD}
\mathcal{M}_m^{(D)}: \sum_{\substack{1\leq r\leq q-1\\r\equiv 1}}\sum_{1\leq s\leq \frac{p-1}{2}}\chi_{r,s}\overline{\chi_{r,s}}+\sum_{\substack{1\leq r\leq q-1\\r\equiv q/2}}\sum_{1\leq s\leq \frac{p-1}{2}}\chi_{r,s}\overline{\chi_{q-r,s}}\period
\eeq
Here $\equiv$ is the equality modulo $2$. On the other hand, the modular $S$-matrix for these representations read (see e.g.~\cite{DiFrancesco:1997nk,Blumenhagen:2009zz}):
\beq
S_{(r,s),(r^{\prime},s^{\prime})}=2(-1)^{1+rs+r^{\prime}s^{\prime}}\sqrt{\frac{2}{p q}}\sin\left(\frac{\pi p}{q}rr^{\prime}\right)\sin\left(\frac{\pi q}{p}ss^{\prime}\right)\period
\eeq
The identity representation corresponds to $(r^{\prime},s^{\prime})=(1,1)$.

As can be seen from \eqref{eq:repA}, all the states in the $A$-series minimal models have identical chiral and anti-chiral representations and all the representations appear only once. Setting $n_{a,a}=1$ in \eqref{eq:generalformulaCFT} and summing all the representations that appear in \eqref{eq:repA}, we obtain the answer that precisely agrees with the result from integrability \eqref{eq:closedform}.
\subsection{Generalization to $D$-series by $\mathbb{Z}_2$-orbifold}\label{sec:dseries}
\paragraph{Derivation.}The staircase model studied above describes the RG flows between diagonal minimal models called the $A$-series. We now propose its simple modification that gives the RG flows between the (non-diagonal) $D$-series minimal models. (The existence of the RG flows connecting two $D$-series minimal models was first pointed out in \cite{Klassen:1991dz}.)

It is well-known in CFT that the $D$-series can be obtained from the $A$-series by gauging the $\mathbb{Z}_2$-symmetry, also known as the $\mathbb{Z}_2$-orbifold, see \cite{DiFrancesco:1997nk}. In 2d QFT, the $\mathbb{Z}_2$-gauging amounts to performing the following manipulations\footnote{Alternatively, we can achieve the gauging by inserting a $\mathbb{Z}_2$ line defect $P$ in all possible ways. This is because the combination $1+P$ inserted along a spatial circle realizes the projection to $\mathbb{Z}_2$-invariant states while $P$ inserted on a time circle gives the twisted sector.} to the original theory:
\begin{enumerate}
\item Add a new sector, called the {\it twisted sector}, in which fields are periodic up to $\mathbb{Z}_2$ transformation; e.g.~$\phi(\sigma+2\pi)=-\phi(\sigma)$.
\item Restrict the Hilbert space to $\mathbb{Z}_2$-invariant states.
\end{enumerate}
The resulting theory possesses an emergent $\mathbb{Z}_2$ symmetry which assigns $+1$ to the untwisted sector and $-1$ to the twisted sector. Gauging this $\mathbb{Z}_2$ takes it back to the original theory \cite{Vafa:1989ih}.

To see the effect of the orbifold at the level of the Bethe ansatz, we need to understand how the $\mathbb{Z}_2$-transformation acts on individual excitations. This is not obvious for the staircase model since it is defined only indirectly as an analytic continuation of the sinh-Gordon model. However, given that the sinh-Gordon model has the $\mathbb{Z}_2$-symmetry which flips the sign of excitations $\phi\to-\phi$, it is natural to assume that the same holds for the staircase model. Then, the $\mathbb{Z}_2$-gauging can be achieved by restricting to states with an even number of particles while allowing them to take either periodic (untwisted sector {\bf S}) or antiperiodic (twisted sector {\bf U}) boundary conditions. 

After imposing the parity-invariance condition needed for the computation of the $p$-function, the Bethe equation for each sector takes
\begin{align}
&{\bf S}:\qquad 1=e^{2ip_jR}S(p_j,-p_j)\prod_{k\neq j}S(p_j,p_k)S(p_j,-p_k)\comma\label{eq:DseriesS}\\
&{\bf U}:\qquad -1=e^{2ip_jR}S(p_j,-p_j)\prod_{k\neq j}S(p_j,p_k)S(p_j,-p_k)\period\label{eq:DseriesU}
\end{align}
We can then proceed as before; namely take the thermodynamic limit and sum over states satisfying the Bethe equations using TBA. The contribution from the {\bf S}-sector was already computed in section \ref{subsec:computation}. The result for the {\bf U}-sector is similar but slightly different since now the Bethe equation {\it does} admit $p_j=0$ as a formal solution. However, since the states in the ${\bf U}$-sector contain an even number of particles, this does not correspond to a physical solution. Thus we subtract it from the final answer using $-\pi \delta(u)$ term as discussed below \eqref{eq:eqforTheta}. The result reads
\beq
|\langle \mathcal{C}|\Omega_L\rangle|^2\quad \overset{{\bf U}}{\supset}\quad \frac{1}{\sqrt{1+Y(0)}}\frac{\det \left[1-\hat{G}_{-}\right]}{\det \left[1-\hat{G}_{+}\right]}\period
\eeq
Summing the two contributions, we get the formula for the $\mathbb{Z}_2$-orbifolded theory
\beq\label{eq:Z2orbifoldfull}
\text{$\mathbb{Z}_2$-orbifold}:\qquad |p|=\left|\langle \mathcal{C}|\Omega_L\rangle\right|=\sqrt{\left(1+\frac{1}{\sqrt{1+Y(0)}}\right)\frac{\det \left[1-\hat{G}_{-}\right]}{\det \left[1-\hat{G}_{+}\right]}}\period
\eeq
Note that, although we have focused on the $p$-function, the argument here can in principle be generalized to other quantities as well, such as the boundary entropies.
\paragraph{Numerical computation.}

\begin{figure}[t]
\centering
\includegraphics[clip, height=6.3cm]{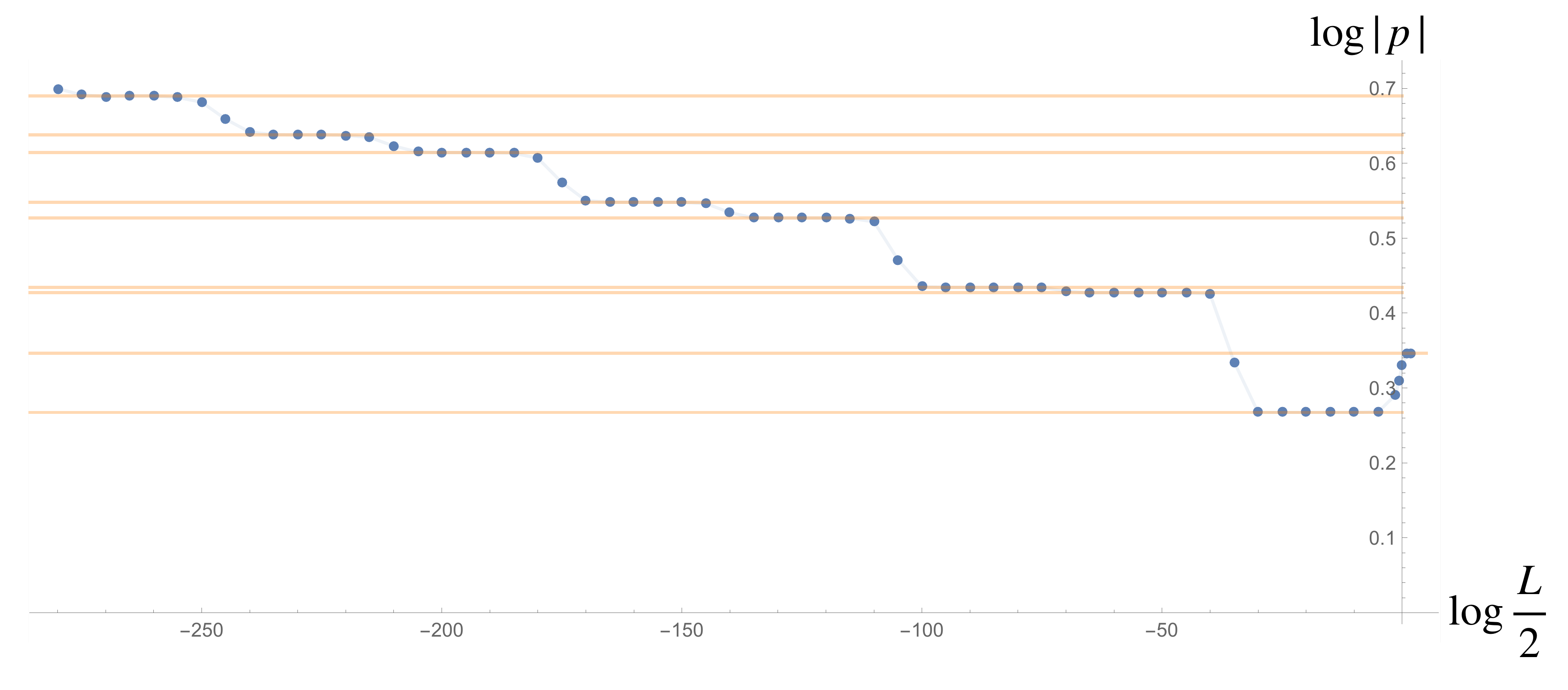}\\
\caption{The $p$-function for the generalized staircase model as a function the volume $L$. Similarly to the effective central charge and the previous example of the $p$-function, this also develops plateaux (orange horizontal lines) at the values of the $p$-function entropy for the minimal models in the $D$-series.}\label{fig:logpd}
\end{figure}

As can be seen in figure \ref{fig:logpd}, the result of the numerical computation exhibits two important differences from that of the non-orbifolded theory. First, although the $p$-function decreases monotonically for most part of the flow, it starts to increase in the deep infrared, after the plateau corresponding to the Ising model. 

Second, the infrared value of the crosscap entropy is $\frac{1}{2}\log2$ while that of the original staircase model is $0$. This difference translates to the following behaviors of the Klein bottle partition function in the infrared:
\beq
Z_{\mathbb{K}}(R,L)\overset{R,L\gg 1}{\sim} \begin{cases}1&\qquad \text{original staircase}\\
2&\qquad \text{orbifolded staircase}\end{cases}\period
\eeq
To understand the physical meaning of this difference, it is useful to analyze $Z_{\mathbb{K}}$ in the loop channel. In the original staircase model, there is a unique vacuum state in the {\bf S}-sector and its contribution dominates in the infrared ($L\gg 1$). This is because all the states in the {\bf T}-sector contain massive excitations. If we further take the infinite $R$ limit, the ground state energy asymptotes to zero and therefore we get $Z_{\mathbb{K}}\sim e^{-LE_{\Omega_{2R}}/2}\sim 1$. 
On the other hand in the orbifolded theory, both of the two sectors ${\bf S}$ and ${\bf U}$ have a state without excitations and their energies both asymptote to zero in the infinite $R$ limit. This is why we get $Z_{\mathbb{K}}\sim 2$. Now, the crucial point is that these two states are oppositely charged under the emergent $\mathbb{Z}_2$-symmetry, which assigns $+1$ to the untwisted sector and $-1$ to the twisted sector. Physically, this means that the emergent $\mathbb{Z}_2$-symmetry is spontaneously broken\footnote{As discussed in Appendix A of \cite{Chai:2020onq}, one can argue more generally that, if the $\mathbb{Z}_2$-symmetry of the original theory is unbroken, the emergent $\mathbb{Z}_2$-symmetry of the orbifolded theory must be broken and vice versa.} in the orbifolded theory in the infrared. 

Although we need to study more examples in order to draw any conclusion, it is tempting to speculate that these two features---the sudden increase of the $p$-function in the infrared and the $\mathbb{Z}_2$-symmetry breaking---might be somehow correlated. As another example where this phenomenon could potentially be observed, it is known that 
the $A$-series minimal models 
flow to a symmetry breaking phase in the deep infrared when perturbed by $\phi_{13}$ with opposite sign, see \cite{Klassen:1991dz} for a related discussion. It would be interesting to study the behavior of the $p$-function also in this case.
Leaving this as an important open question, let us emphasize here that, even for the orbifolded model, the $p$-function keeps decreasing until the deep infrared. In particular, it means that the $p$-function monotically decreases along the flows that connect neighboring $D$-series minimal models.
\paragraph{Analytic expressions and comparison with CFT.} As in the $A$-series, we can compute the values at plateaux analytically by simply plugging in \eqref{eq:plateauratio} and \eqref{eq:plateauy0} into \eqref{eq:Z2orbifoldfull}. The result reads
\beq\label{eq:closedformD}
\mathcal{M}^{(D)}_m:\quad |p|=|\langle \mathcal{C}|\Omega\rangle|=\begin{cases}\left(\frac{2}{m(m+1)}\right)^{\frac{1}{4}}\sqrt{\left(\cot\frac{\pi}{2m}\right)\left(1+\frac{1}{\sin \frac{\pi}{m+1}}\right)}\qquad &m:\text{ odd}\\\left(\frac{2}{m(m+1)}\right)^{\frac{1}{4}}\sqrt{\left(\cot\frac{\pi}{2(m+1)}\right)\left(1+\frac{1}{\sin \frac{\pi}{m}}\right)}\qquad &m:\text{ even}\end{cases}\period
\eeq
The explicit results for the first few are
\beq
\begin{aligned}
&\mathcal{M}^{(D)}_3 :\sqrt{1+\frac{1}{\sqrt{2}}}\comma\qquad&&\mathcal{M}^{(D)}_{4}:\sqrt{1+\frac{1}{\sqrt{2}}}\left(1+\frac{2}{\sqrt{5}}\right)^{\frac{1}{4}}\comma\\
&\mathcal{M}^{(D)}_{5}:\left(3+\frac{6}{\sqrt{5}}\right)^{\frac{1}{4}}\period
\end{aligned}
\eeq
The results for $\mathcal{M}_{3,4}^{(D)}$ coincide with those for the $A$-series. This is expected since the $A$- and $D$-series minimal models are identical for $m=3,4$. The result for $m=5$ is different from that of the $A$-series and corresponds to the 3-state Potts model.

The closed-form expression for the $D$-series \eqref{eq:closedformD} is in perfect agreement with the direct CFT computation which can be obtained by using the operator content given in \eqref{eq:repD} and the formula\footnote{Note however that we are making an assumption in the computation that the scalar primaries have eigenvalues $+1$. See a paragraph below \eqref{eq:generalformulaCFT}.} \eqref{eq:generalformulaCFT}. This gives an important cross-check of our integrability formula \eqref{eq:Z2orbifoldfull} for the $\mathbb{Z}_2$-orbifolded model.

\subsection{Fermionization of integrable field theories} \label{sec:fermi}
For every two-dimensional quantum field theory with non-anomalous $\mathbb{Z}_2$ symmetry, one can define four different versions by a combination of $\mathbb{Z}_2$-orbifolding and tensoring with the {\it Arf theory}---a spin TQFT which describes the low energy limit of the Kitaev chain \cite{Kitaev:2000nmw} and whose partition function is given by the Arf invariant $(-1)^{{\rm Arf}[\rho]}$. This latter manipulation is called {\it fermionization}, and for minimal models it was studied by Petkova already in the eighties, see \cite{Petkova:1988cy,Furlan:1989ra}. More recently, it has been attracting interest \cite{Karch:2019lnn,Runkel:2020zgg,Hsieh:2020uwb,Kulp:2020iet,Smith:2021luc,Fukusumi:2021zme,Ebisu:2021acm} in the context of fermionic Symmetry Protected Topological phases. Interestingly, a similar concept was proposed in the context of integrable field theories by Klassen and Melzer \cite{Klassen:1992eq}, who pointed out that the $S$-matrices of the sine-Gordon and massive Thirring models differ by a sign. The aim of this subsection is to discuss the relation between the two concepts. A more detailed analysis will be presented elsewhere.

Following the notations in \cite{Hsieh:2020uwb}, we call the four variants $A$, $D$, $F$ and $\tilde{F}$: $A$ is the original theory with $\mathbb{Z}_2$ symmetry while $D$ is its $\mathbb{Z}_2$-orbifold. $F$ and $\tilde{F}$ are both fermonic and can be obtained from $A$ and $D$ by fermionization. Since $F$ and $\tilde{F}$ differ only by the assignment of the fermion number in the Ramond (R) sector (=periodic sector), we focus on $A$, $D$ and $F$ in what follows. 

\begin{table}[t]
\centering
\begin{tabular}{c|cc}
$A$-theory&$I$&$P$\\\hline
$\mathbb{Z}_2$-even&{\bf S}&{\bf U}\\
$\mathbb{Z}_2$-odd&{\bf T}&{\bf V}
\end{tabular}
\hspace{50pt}
\begin{tabular}{c|cc}
$D$-theory&$I$&$P$\\\hline
$\mathbb{Z}_2$-even&{\bf S}&{\bf T}\\
$\mathbb{Z}_2$-odd&{\bf U}&{\bf V}
\end{tabular}
\hspace{50pt}
\begin{tabular}{c|cc}
$F$-theory&NS&R\\\hline
bosonic&{\bf S}&{\bf U}\\
fermionic&{\bf V}&{\bf T}
\end{tabular}
\caption{Different sectors in $A$, $D$ and $F$-theories (adapted from \cite{Hsieh:2020uwb}). $I$ and $P$ correspond to the standard spectrum and the defect spectrum respectively. As usual, the NS- and R-sectors correspond to antiperiodic and periodic boundary conditions.}\label{tab:table}
\end{table}

As explained in \cite{Hsieh:2020uwb} and summarized in table \ref{tab:table}, each theory consists of the same four sectors {\bf S}, {\bf T}, {\bf U} and {\bf V}, but their interpretation depends on the theory we consider: The left columns of the $A$- and $D$-theories in table \ref{tab:table} give the spectrum in the standard Hilbert space while the right columns give the ``defect spectrum''; the spectrum at the endpoint of the $\mathbb{Z}_2$-defect line $P$. The rows in $A$ and $D$ distinguish states with different $\mathbb{Z}_2$-parity. On the other hand, the four sectors in the $F$-theory are classified by the periodicity and the fermion number. For instance, the {\bf S}-sector corresponds to $\mathbb{Z}_2$-even states in the standard Hilbert space of the $A$-theory while it corresponds to the bosonic Neveu-Schwarz (NS) sector (=anti-periodic sector) in the $F$-theory.

Let us now see how fermionization is realized at the level of the Bethe equations. We begin with the Bethe equations in the $A$-theory. We already wrote down three of them when the parity condition is imposed (see \eqref{eq:Sparbethe}, \eqref{eq:Tparbethe} and \eqref{eq:DseriesU}), but we now present them all without imposing the parity condition:
\beq
\begin{aligned}
&{\bf S}:\qquad 1=e^{ip_j R}\prod_{k\neq j}^{2M}S(p_j,p_k)\comma\qquad &&{\bf U}:\qquad -1=e^{ip_j R}\prod_{k\neq j}^{2M}S(p_j,p_k)\comma\\
&{\bf T}:\qquad 1=e^{ip_j R}\prod_{k\neq j}^{2M+1}S(p_j,p_k)\comma\qquad &&{\bf V}:\qquad -1=e^{ip_j R}\prod_{k\neq j}^{2M+1}S(p_j,p_k)\period
\end{aligned}
\eeq
Here we set the size of the system to be $R$. We next rewrite them into the Bethe equations in the $F$-theory. The proposal made by Klassen and Melzer is that the $S$-matrix of a fermionic theory can be obtained by flipping the sign\footnote{This is roughly because the $\mathbb{Z}_2$-odd excitation in the original theory comes attached to a $\mathbb{Z}_2$ Wilson line after fermionization and that gives an extra $-1$ factor upon exchanging particles.} of the $S$-matrix of a bosonic theory, 
\beq\label{eq:fermionSmatrix}
S_{f}(p,q)=-S(p,q)\period
\eeq
 Rewriting each of these equations in terms of $S_f$, we obtain the following set of equations:
\beq\label{eq:fermionicBetheeq}
\begin{aligned}
&{\bf S}:\qquad -1=e^{ip_j R}\prod_{k\neq j}^{2M}S_{f}(p_j,p_k)\comma\qquad &&{\bf U}:\qquad 1=e^{ip_j R}\prod_{k\neq j}^{2M}S_{f}(p_j,p_k)\comma\\
&{\bf T}:\qquad 1=e^{ip_j R}\prod_{k\neq j}^{2M+1}S_{f}(p_j,p_k)\comma\qquad &&{\bf V}:\qquad -1=e^{ip_j R}\prod_{k\neq j}^{2M+1}S_{f}(p_j,p_k)\period
\end{aligned}
\eeq
Note that the signs of the left hand side of {\bf S} and {\bf U} are flipped since they contain an odd number of $S$-matrices. Now, we interpret these equations in terms of the $F$-theory. The left hand sides of {\bf S} and {\bf V} are $-1$, indicating that they are in the NS-sector, while {\bf U} and {\bf T} are in the R-sector. The fermion number of each sector can be determined simply by counting the number of excitations. As a result, we conclude
\beq
\begin{aligned}
&{\bf S}:\quad \text{bosonic NS-sector}\comma\qquad &&{\bf U}:\quad \text{bosonic R-sector}\comma\\
&{\bf T}:\quad \text{fermionic R-sector}\comma\qquad &&{\bf V}:\quad \text{fermionic NS-sector}\period
\end{aligned}
\eeq
Happily, this agrees precisely with the results in \cite{Hsieh:2020uwb} summarized in table \ref{tab:table}, supporting the proposal by Klassen and Melzer. To recover the Hilbert space of the $A$- or $D$-theories, one simply needs to pick two of the four sectors by correlating the number of excitations with the periodicity. This is basically the GSO projection in superstring theory \cite{Gliozzi:1976jf,Gliozzi:1976qd,Seiberg:1986by}.

In summary, our conclusion is
\begin{enumerate}
\item Fermionization of the Bethe equation is achieved simply by flipping the sign of the $S$-matrix $S_f=-S$ and allowing twisted boundary conditions.
\item The GSO projection of the Bethe equation amounts to picking two of the four sectors in the fermionic theory by correlating the number of excitations with the periodicity.
\end{enumerate}
Note that the discussion above is focused on the torus partition function, and 
it would be interesting to generalize it to non-orientable manifolds. That would allow us to study integrable worldsheet theories for unoriented type 0 string (see \cite{Skrzypek:2021eue} for a related discussion), for which the existence of eight different GSO projections was pointed out recently \cite{Kaidi:2019pzj,Kaidi:2019tyf}.
\paragraph{Clarifying remarks.} Let us make an additional remark on fermionization of the $S$-matrix \eqref{eq:fermionSmatrix}. A similar change of the sign is often discussed in the literature in the discussion of the so-called ``bosonic TBA'' \cite{Mussardo:1999aj,Cordova:2021fnr}.  In this context, one starts with a standard $S$-matrix of integrable theories and flips its sign. The resulting $S$-matrix satisfies $S(0)=+1$, indicating that particles with the same mode number are now allowed to coexist. This leads to a substantial change of the TBA equation (for instance $\log (1+Y)$ changes to $-\log (1-Y)$), and it typically leads to a Hagedorn-like instability at finite volume \cite{Mussardo:1999aj,Cordova:2021fnr}. 

We emphasize that this notion of ``bosonic'' and ``fermionic'' is different from what has been discussed here: In the context of ``bosonic TBA'', the notion of ``bosonic'' or ``fermionic'' refers to whether or not the exclusion principle is realized. In contrast, in this paper we followed a standard convention in quantum field theory, in which fermionic theories refer to theories that depend on the spin structure while bosonic theories are the ones that are insensitive to it. In order to make the distinction clear, we call fermionic theories discussed in our paper {\it spin theories} and bosonic theories {\it non-spin theories} in the discussion below. 

The difference of the two notions can be seen clearly in the following fact. As pointed out by Klassen and Melzer \cite{Klassen:1992eq}, for the exclusion principle to be satisfied, spin theories need to have $S(0)=+1$ while non-spin theories need to have $S(0)=-1$. Despite this difference of $S(0)$, both are described by the standard TBA (``fermionic TBA''). More generally, the relation between the two concepts can be summarized as follows:
\beq
\begin{tabular}{c|cc}
&$S(0)=-1$&$S(0)=+1$\\\hline
\text{non-spin}&\text{fermionic TBA}&\text{bosonic TBA}\\
\text{spin}&\text{bosonic TBA}&\text{fermionic TBA}
\end{tabular}
\eeq
In this paper, we discussed the diagonal entries in this table and how to transform one to the other. In contrast the literature on ``bosonic TBA'' addresses theories on the first row.
\section{Crosscap States in Spin Chain\label{sec:spin}}
We now construct analogs of crosscap states in integrable spin chains. We focus on two simple rank-$1$ spin chains, the XXX spin chain and the noncompact $SL(2,R)$ spin chain.
\subsection{Definition and entanglement structures\label{subsec:defent}}
\paragraph{XXX spin chain.} Let us first discuss the simplest integrable spin chain; the XXX (Heisenberg) spin chain. At each site, we have the spin $1/2$ representation of $SU(2)$ and the local Hilbert space is spanned by $\{|\!\!\uparrow\rangle, |\!\!\downarrow\rangle\}$. It has a nearest-neighbor Hamiltonian invariant under the global $SU(2)$ transformation,
\beq\label{eq:su2hamil}
H_{\text{SU(2)}}\propto \sum_j\vec{S}_j\vec{S}_{j+1}\period
\eeq

The Hamiltonian \eqref{eq:su2hamil} is known to be solvable using the Bethe ansatz and the energy eigenstate (called the Bethe state) is given by
\beq
|{\bf u}\rangle=\sum_{n_1<\cdots <n_M}\Psi_{\bf u}(n_1,\ldots, n_M)|\!\uparrow \cdots \underset{n_1}{\!\!\downarrow}\underset{\cdots}{\cdots}\underset{n_M}{\!\!\downarrow}\cdots \!\!\uparrow\rangle\comma
\eeq
where the wave function is given by
\beq
\Psi_{\bf u}(n_1,\ldots, n_M)=\left(\prod_{j<k}^{M}h_{\rm SU}(u_j,u_k)\right)\sum_{\sigma\in S_M}\prod_{j<k}^{M}\frac{1}{h_{\rm SU}(u_{\sigma_j},u_{\sigma_k})}\prod_{j=1}^{M}e^{ip(u_{\sigma_j})n_j}\comma
\eeq
with
\beq
e^{ip(u)}\equiv\frac{u+\frac{i}{2}}{u-\frac{i}{2}}\comma \qquad h_{\rm SU}(u,v)\equiv \frac{u-v}{u-v+i}\period 
\eeq
The rapidity set ${\bf u}$ needs to satisfy the Bethe equation
\beq
1=e^{ip_jL}\prod_{k\neq j}S_{\rm SU}(u_j,u_k)\comma
\eeq
where $S_{\rm SU}(u,v)$ is the $S$-matrix given by $S_{\rm SU}(u,v)=h_{\rm SU}(u,v)/h_{\rm SU}(v,u)$ and $L$ is the length of the chain. For details, see standard reviews and textbooks (for instance \cite{Faddeev:1996iy,Korepin:1993kvr}).

To define the crosscap state in this model, we simply mimic the definition in the field theory; namely we identify states at antipodal sites. To be more concrete, we first take an entangled pair of states at the $j$-th and $(j+\frac{L}{2})$-th sites
\beq\label{eq:defsmallc}
|c\rangle\!\rangle_{j}\equiv |\!\!\uparrow\rangle_j\otimes|\!\!\uparrow\rangle_{j+\frac{L}{2}}+|\!\!\downarrow\rangle_j\otimes|\!\!\downarrow\rangle_{j+\frac{L}{2}}\comma
\eeq
where $|\bullet\rangle_j$ means a state at the $j$-th site. We then consider a tensor product of such states over the chain,
\beq
|\mathcal{C}\rangle\equiv \prod_{j=1}^{\frac{L}{2}}\left(|c\rangle\!\rangle_j\right)^{\otimes}\period
\eeq
This gives a straightforward analog of the crosscap state in the XXX spin chain. We will later see that the overlap with the Bethe state is given by a ratio of determinants.
\paragraph{SL(2,R) spin chain.} The crosscap state can be defined in the non-compact $SL(2,R)$ spin chain as well. In this case, the spin chain is defined in terms of the  $SL(2,R)$ algebra
\beq
[S_0,S_{\pm}]=\pm S_{\pm}\comma\qquad [S_{+},S_{-}]=-2S_0\comma
\eeq
and the Hilbert space at each site is spanned by\footnote{In the application to $\mathcal{N}=4$ SYM, the generator $S_{+}$ is identified with a covariant derivative along a lightcone direction $D_{+}$.}
\beq\label{eq:bethesl2}
|n\rangle\equiv \frac{(S_{+})^{n}}{n!}|0\rangle\comma\qquad n\in \mathbb{Z}_{>0}\comma
\eeq
with the vacuum state $|0\rangle$ defined by $S_{-}|0\rangle=0$. For details, see \cite{Beisert:2004ry,Braun:1998id,Braun:1999te,Belitsky:1999qh,Belitsky:1999ru}.

The Bethe state of this spin chain is given by
\beq
|{\bf u}\rangle=\sum_{n_1\leq\cdots \leq n_M}\Psi_{\bf u}(n_1,\ldots, n_M)|0 \cdots \underset{n_1}{1}\underset{\cdots}{\cdots}\underset{n_M}{1}\cdots 0\rangle\comma
\eeq
with
\beq
\tilde{\Psi}_{\bf u}(n_1,\ldots, n_M)=\left(\prod_{j<k}^{M}h_{\rm SL}(u_j,u_k)\right)\sum_{\sigma\in S_M}\prod_{j<k}^{M}\frac{1}{h_{\rm SL}(u_{\sigma_j},u_{\sigma_k})}\prod_{j=1}^{M}e^{ip(u_{\sigma_j})n_j}\comma
\eeq
and
\beq
h_{\rm SL}(u,v)\equiv \frac{u-v}{u-v-i}\period 
\eeq
The Bethe equation for the $SL(2,R)$ chain is given by
\beq
1=e^{ip_jL}\prod_{k\neq j}S_{\rm SL}(u_j,u_k)\comma
\eeq
with $S_{\rm SL}(u,v)=h_{\rm SL}(u,v)/h_{\rm SL}(v,u)$.
Note that, unlike the XXX spin chain, several excitations can be at the same site. For instance, when $n_1=n_2$ in the sum in \eqref{eq:bethesl2}, we will have $|0\cdots \underset{n_1=n_2}{2}\cdots \rangle$ as a ket.

\begin{figure}[t]
\centering
\begin{minipage}{0.49\hsize}
\centering
\hspace{-50pt}\includegraphics[clip, height=3.3cm]{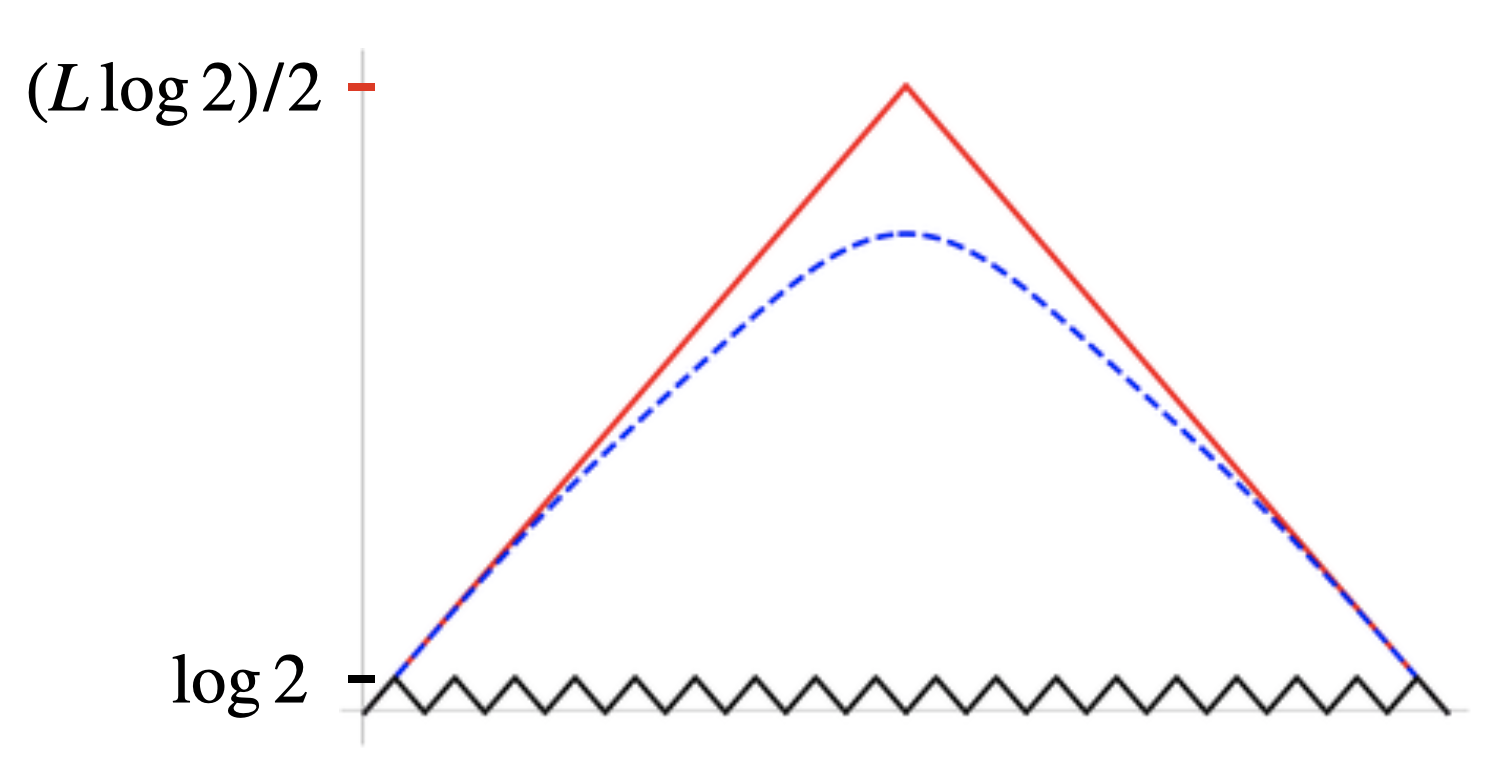}\\
(a)
\end{minipage}
\begin{minipage}{0.49\hsize}
\centering
\includegraphics[clip, height=3.3cm]{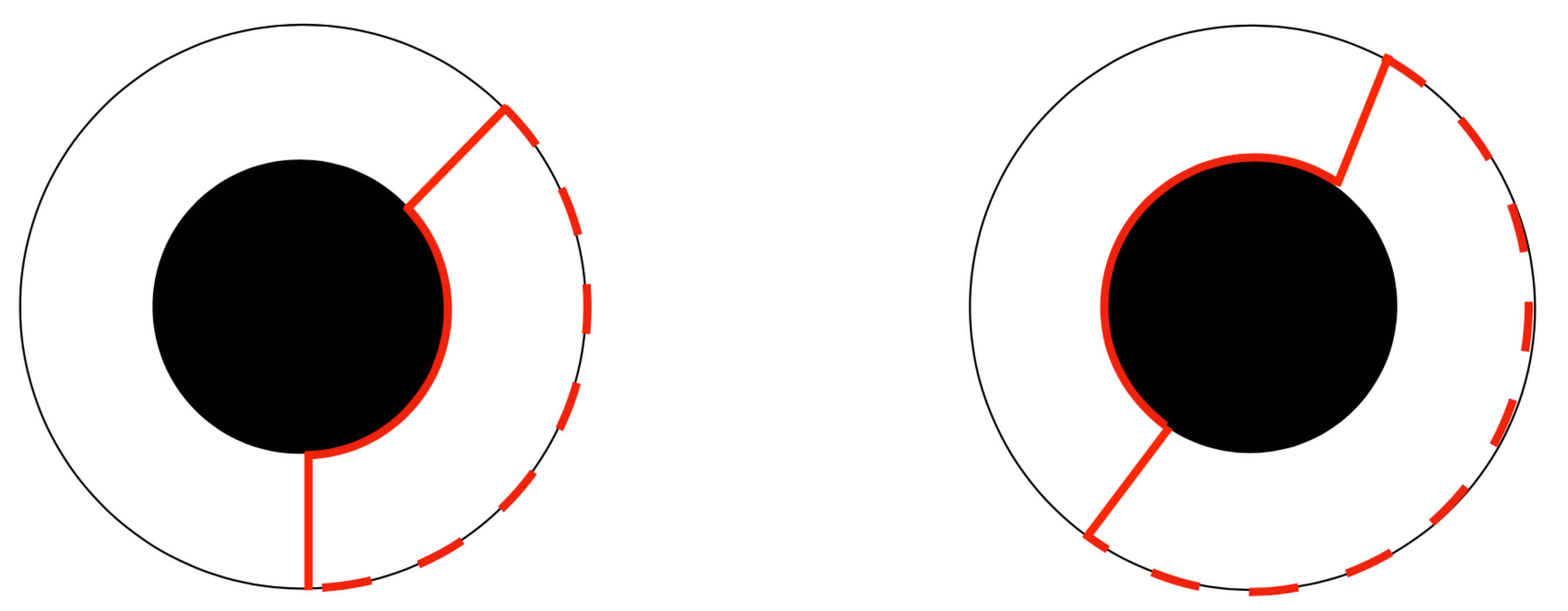}\\
(b)
\end{minipage}
\caption{(a) The subregion entanglement entropy for a crosscap state (red curve), a generic chaotic eigenstate (dashed curve), and a boundary state (black curve). (b) The phase transition of the Ryu-Takayanagi (RT) surface. The outer circle is the spatial slice of the boundary CFT while the black dot in the middle is a black hole. The area of the RT surface (denoted by a solid red curve) grows linearly until the subregion of the boundary CFT (denoted by a dashed red curve) exceeds half the system size. After that, the RT surface discontinuously jumps from the one on the right to the one on the left and starts decreasing linearly.}\label{fig:entangle}
\end{figure}

The crosscap state in this model is defined by
\beq
|\mathcal{C}\rangle\equiv\prod_{j=1}^{\frac{L}{2}}\left(|\tilde{c}\rangle\!\rangle_j\right)^{\otimes}\comma
\eeq
where $|\tilde{c}\rangle\!\rangle_j$ is the antipodally identified two-site state
\beq
|\tilde{c}\rangle\!\rangle_j\equiv \sum_{n=0}^{\infty}|n\rangle_j\otimes|n\rangle_{j+\frac{L}{2}}\period
\eeq
\paragraph{Entanglement structures.} The boundary states in the XXX spin chain can also be expressed in terms of entangled two-site states \cite{Piroli:2017sei}. However, while the crosscap states in the XXX chain are given by two-site states at antipodal sites, the boundary states are given by two-site states at neighboring sites. This difference is reflected in the different entanglement structures: the boundary state is short-range entangled, and their subregion entanglement entropy never exceeds $\log 2$ (see figure \ref{fig:entangle}-(a)). This is simply because the boundary states are tensor products of local two-site states. 

In contrast, the subregion entanglement entropy of the crosscap state exhibits the volume law: it increases linearly until the size of the region reaches half of the system size and the entanglement entropy takes the maximum value $\frac{L}{2}\log 2$. Thereafter, it decreases linearly until it goes back to zero (see figure \ref{fig:entangle}-(a)). The linear growth of the entanglement entropy is a generic feature of high energy eigenstates in chaotic spin chains. However, in general, the linear growth stops before the subregion size reachs half of the system size and the maximum value of the entanglement entropy is strictly smaller than $\frac{L}{2}\log 2$, see \cite{Nakagawa:2017yiw,Vidmar:2017pak,Murthy:2019qvb,Dong:2020iod}.

An example in which the linear growth continues to half the system size is a pure-state black hole in AdS$_3$, which is holographically dual to a high energy eigenstate in 2d CFT with infinite central charge. In this example, one can calculate the entanglement entropy using the Ryu-Takayanagi formula \cite{Ryu:2006bv} and show that the entropy grows linearly until the subregion covers half of the system. After that point it makes a sharp turn and decreases linearly due to the phase transition of the Ryu-Tanayanagi surface, see figure \ref{fig:entangle}-(b).

This similarity motivates to use our crosscap states as a toy model of pure-state black holes\footnote{See \cite{Maloney:2016gsg,Tsiares:2020ewp,Verlinde:2015qfa,Nakayama:2016xvw} for other recent discussions on the crosscap states in the context of holography.}. Of course, since the spin chains we are studying are integrable while the black holes are expected to be dual to chaotic quantum systems, we cannot hope for precise correspondences between the dynamics of the two systems. Nevertheless we may be able to learn some aspects. For instance, the question  ``How do pure-state black holes evaporate?'' can be translated to the question ``How do highly-entangled crosscap states decohere as a result of dynamics?''. To address this  question, we need a system in which the entanglement stored in the correlation of spins can leak into other dynamical degrees of freedom. There are several ways to achieve this but one possible approach is to consider an integrable system with both spins and particles as dynamical degrees of freedom such as the spin Calogero-Sutherland model see e.g.~\cite{Minahan:1992ht}. It would be interesting to construct the crosscap states in such models and use them as initial conditions of quantum quench.

Even without the connection to the blackhole physics, our crosscap states provide interesting initial conditions for quantum quench which are analytically tractable but are quite distinct from the ones given by integrable boundary states. Thus it is certainly worth studying the details of their dynamics. We leave these exciting questions to future investigations.
\subsection{Determinant formulae for overlaps}
With the crosscap states we constructed, we computed overlaps with various Bethe states. As a result, we found a selection rule identical to that of the integrable boundary state: namely the overlap vanishes unless the set of rapidities are parity invariant,
\beq\label{eq:selectionrule}
{\bf u}=\{u_1,\ldots, u_M\}\qquad \qquad u_{j+\frac{M}{2}}=-u_{j}\period
\eeq
In addition, we found convincing evidence that the overlap is given by the following universal formula, both in the XXX spin chain and the $SL(2,R)$ spin chain:
\beq\label{eq:overlapspin}
\langle \mathcal{C}|{\bf u}\rangle=\left(\prod_{1\leq j<k\leq M}S(u_j,u_k)\right)^{\frac{1}{2}}\left(\prod_{j=1}^{M}\frac{1}{\partial_{u_j}p (u_j)}\right)^{\frac{1}{2}}\det G_{+}\period
\eeq
Here $S(u,v)$ is the $S$-matrix, which is taken to be $S=S_{\rm SU}$ for the XXX chain and $S=S_{\rm SL}$ for the $SL(2,R)$ chain. The matrices $G_{\pm}$ are given by \eqref{eq:gpm0}, which we display again for convenience:
\beq
\begin{aligned}
\left(G_{\pm}\right)_{1\leq i,j\leq\frac{M}{2}}&=\left[L\del_{u}p(u_i)+\sum_{k=1}^{\frac{M}{2}}\mathcal{K}_{+}(u_i,u_k)\right]\delta_{ij}-\mathcal{K}_{\pm}(u_i,u_j)\comma\\
\mathcal{K}_{\pm}(u,v)&=\frac{1}{i}\del_u\left[\log S(u,v)\pm \log S(u,-v)\right]\period
\end{aligned}
\eeq

To compare this expression with the results in section \ref{sec:field}, we need to normalize the state $|{\bf u}\rangle$. This can be achieved using the result for the norm\footnote{Here we defined the norm using the two-point function in $\mathcal{N}=4$ SYM. This is slightly different from the standard spin-chain norm since it does not involve the complex conjugation of the wave functions. For more details, see Appendix C of \cite{Jiang:2019xdz}.} of parity-symmetric states \cite{deLeeuw:2015hxa,Jiang:2019xdz}
\beq
\langle {\bf u}|{\bf u}\rangle=\left(\prod_{1\leq j<k\leq M}S(u_j,u_k)\right)\left(\prod_{j=1}^{M}\frac{1}{\partial_{u_j}p (u_j)}\right)\det G_{+}\det G_{-}\period
\eeq
We then arrive at an exceedingly simple expression, which is identical to the asymptotic limit of the crosscap overlap \eqref{eq:asymptotic} obtained in section \ref{sec:field}:
\beq\label{eq:spinnormalized}
\begin{aligned}
\frac{\langle \mathcal{C}|{\bf u}\rangle}{\sqrt{\langle {\bf u}|{\bf u}\rangle}}=\sqrt{\frac{\det G_{+}}{\det G_{-}}}\period
\end{aligned}
\eeq
The equivalence between \eqref{eq:asymptotic} and \eqref{eq:spinnormalized} supports our interpretation of $|\mathcal{C}\rangle$ as the spin-chain analog of the crosscap states. It is also plausible that the two are more directly related in $\mathcal{N}=4$ SYM; namely, there might be some observable in $\mathcal{N}=4$ SYM which are given by the spin-chain crosscap state at weak coupling and the crosscap state on the integrable worldsheet theory at finite coupling. See also the discussion in section \ref{sec:conclusion}.
\subsection{Integrability of the crosscap state}
The selection rule \eqref{eq:selectionrule} suggests that the crosscap states $|\mathcal{C}\rangle$ are annihilated by infinitely many odd conserved charges. Below we will prove this explicitly for the XXX spin chain.

Let us first recall the Lax operator of the XXX spin chain
\beq
L_j(u)\equiv \pmatrix{cc}{u+iS^{z}_{j}&i S^{-}_{j}\\S^{+}_j&u-iS^{z}_{j}}\comma
\eeq
where $S_j^{\pm, z}$ are the generators of SU(2) acting on the spin Hilbert space at the $j$-th site. Using the Lax operator, we can define the transfer matrix $T(u)$, which generates higher conserved charges:
\beq
T(u)\equiv {\rm tr}_{V}\left(L_1(u)\cdots L_L(u)\right)\period
\eeq
Here $V$ denotes the auxiliary space, not the spin-chain Hilbert space. In what follows, we will show that the crosscap state satisfies
\beq
\langle \mathcal{C}|T (u)=\langle \mathcal{C}|T (-u)\period
\eeq
This equality means that the crosscap state is annihilated by infintely many odd conserved charges generated by $T(u)-T(-u)$.

\begin{figure}[t]
\centering
\includegraphics[clip, height=3cm]{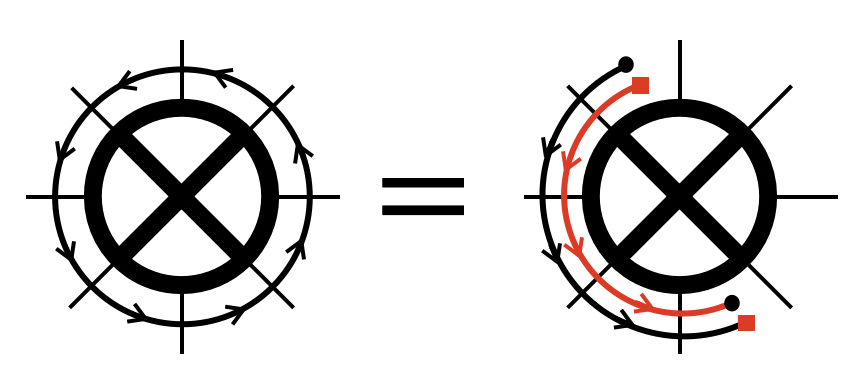}
\caption{Using the property \eqref{eq:basicrewriting}, one can rewrite the action of the transfer matrix $T(u)$ on the crosscap state to an operator which only acts on a half of the spin chain. See \eqref{eq:rewriting0}.}\label{fig:spin1}
\end{figure}

As the first step, we consider the following equality\footnote{The derivation below is partially motivated by the discussion on the monodromy relation for the three-point function in $\mathcal{N}=4$ SYM \cite{Kazama:2014sxa,Jiang:2014cya} and the proof of integrability of boundary states in spin chains in \cite{Piroli:2017sei,Jiang:2020sdw}.}
\beq\label{eq:basicrewriting}
{}_j\langle\!\langle c|L_{j+\frac{L}{2}}(u)=-{}_j\langle\!\langle c|\left(\sigma_2L_{j}(-u)\sigma_2\right)\comma
\eeq
where $\sigma_2$ is a Pauli matrix acting on the space $V$. This equality can be verified straightforwardly by computing both sides using the definition of $|c\rangle\!\rangle$ \eqref{eq:defsmallc}. Using this relation, one can rewrite the action of the transfer matrix as follows, see also figure \ref{fig:spin1} :
\beq\label{eq:rewriting0}
\begin{aligned}
\langle \mathcal{C}|T(u)&=\langle \mathcal{C}|\left(L_1(u)\cdots L_{\frac{L}{2}}(u)\right)_{ab}\left(L_{\frac{L}{2}+1}(u)\cdots L_{L}(u)\right)_{ba}\\
&=(-1)^{\frac{L}{2}}\langle \mathcal{C}|\left(\sigma_2 L_{1}(-u)\cdots L_{\frac{L}{2}}(-u)\sigma_2\right)_{ba}\left(L_1(u)\cdots L_{\frac{L}{2}}(u)\right)_{ab}\\
&=(-1)^{\frac{L}{2}}\langle \mathcal{C}|\left(\sigma_2 L_{1}(-u)\cdots L_{\frac{L}{2}}(-u)\right)_{ba}\left(\sigma_2 L_1(u)\cdots L_{\frac{L}{2}}(u)\right)_{ab}\period
\end{aligned}
\eeq
Here $a$ and $b$ are matrix indices which are summed over.

\begin{figure}[t]
\centering
\includegraphics[clip, height=5.5cm]{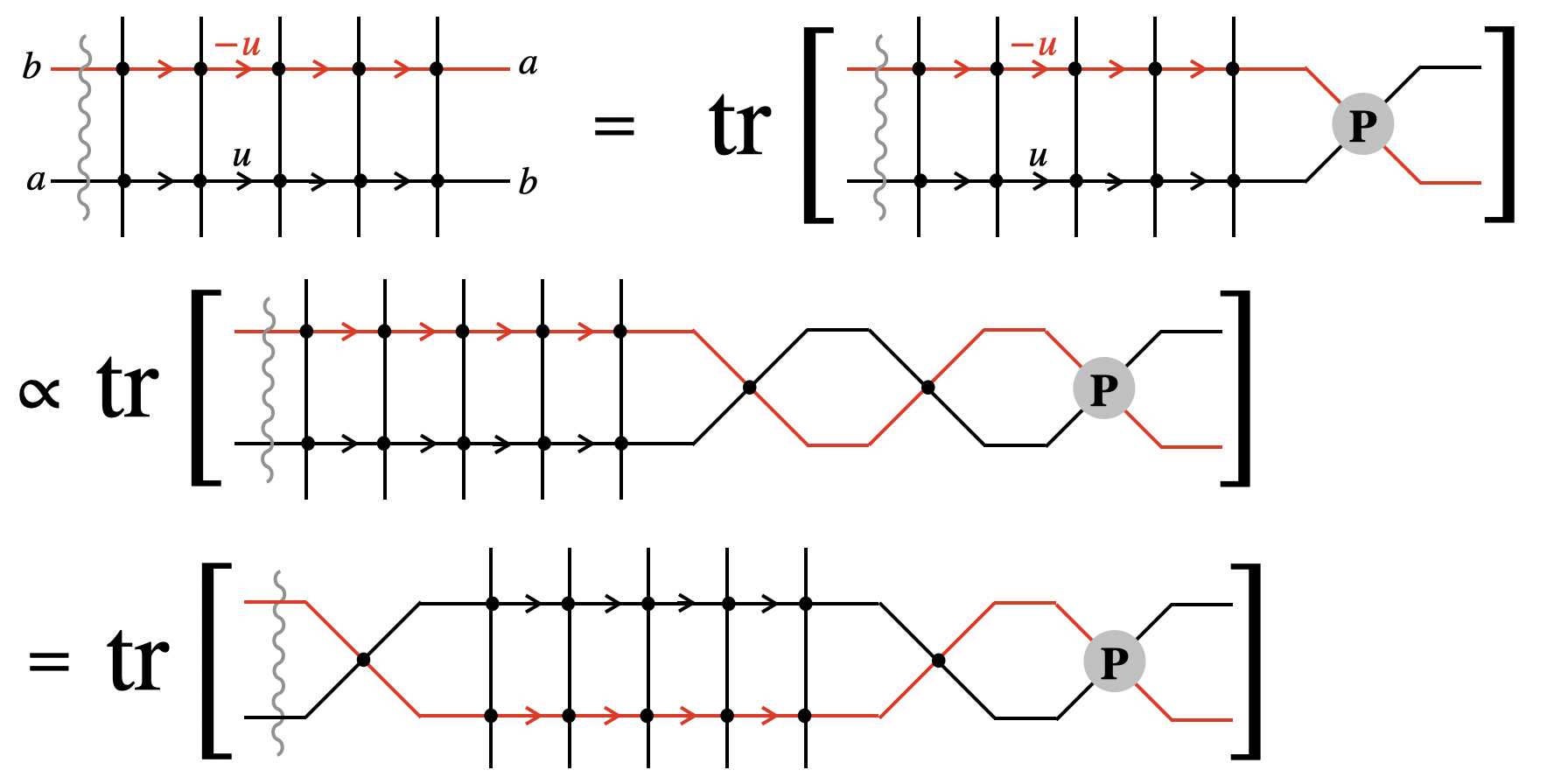}
\caption{Proof of integrability of the crosscap state. The wavy gray curve signifies the action of $\sigma_2$, the gray circle with ${\bf P}$ inside is the permutation operator, and the black dots are the $R$-matrices. On the second line, we used the identity satisfied by the $R$-matrices \eqref{eq:R-mat} while on the third line  we used the Yang-Baxter relation to move the $R$-matrix.}\label{fig:spin2}
\end{figure}

As the next step, we rewrite it as
\beq
\text{\eqref{eq:rewriting0}}=(-1)^{\frac{L}{2}}\langle \mathcal{C}|\left(\sigma_2 L_{1}(-u)\cdots L_{\frac{L}{2}}(-u)\right)_{ab}\left(\sigma_2 L_1(u)\cdots L_{\frac{L}{2}}(u)\right)_{\bar{a}\bar{b}} \delta_{\bar{b}a}\delta_{b\bar{a}}\period
\eeq
This is just a trivial rewriting but now we can interpret the barred indices $\bar{a}$ and $\bar{b}$ as acting on another auxiliary space $\bar{V}$, and identify $\delta_{\bar{b}a}\delta_{b\bar{a}}$ with the permutation operator ${\bf P}$ acting on $V\otimes \bar{V}$. This leads to the following expression (see also figure \ref{fig:spin2})
\beq\label{eq:rewriting1}
\text{\eqref{eq:rewriting0}}=(-1)^{\frac{L}{2}}\langle \mathcal{C}|{\rm tr}_{V\otimes \bar{V}}\left[(\sigma_2\otimes \sigma_2)(L_1(-u)\otimes L_1(u))\cdots(L_{\frac{L}{2}}(-u)\otimes L_{\frac{L}{2}}(u)){\bf P}\right]\period
\eeq

The third step is to consider the $R$-matrix acting on $V\otimes \bar{V}$,
\beq
R(u)=u{\bf I}+i{\bf P}\comma
\eeq
with ${\bf I}$ being the identity operator, and use the equality
\beq\label{eq:R-mat}
R(-2u)R(2u)=-(4u^2+1)\period
\eeq
Inserting this equality inside \eqref{eq:rewriting1}, we get
\begin{align}
&\langle \mathcal{C}|T(u)=\frac{(-1)^{\frac{L}{2}+1}}{(4u^2+1)}\label{eq:rewriting2}\\
&\times \langle \mathcal{C}|{\rm tr}_{V\otimes \bar{V}}\left[(\sigma_2\otimes \sigma_2)(L_1(-u)\otimes L_1(u))\cdots(L_{\frac{L}{2}}(-u)\otimes L_{\frac{L}{2}}(u))R(-2u)R(2u){\bf P}\right]\period\nonumber
\end{align}

The last step is to move the $R$-matrix using the Yang-Baxter relation (RLL relation):
\beq\label{eq:YB}
\left(L_{j}(-u)\otimes L_j(u)\right)R(-2u)=R(-2u)\left(L_{j}(u)\otimes L_j(-u)\right)\period
\eeq 
Since both $(\sigma_2\otimes \sigma_2)$ and ${\bf P}$ commute with the $R$-matrix, we can rewrite \eqref{eq:rewriting2} into the following by repeated use of \eqref{eq:YB} (see figure \ref{fig:spin2}):
\begin{align}
&\langle \mathcal{C}|T(u)=\frac{(-1)^{\frac{L}{2}+1}}{(4u^2+1)}\label{eq:rewriting3}\\
&\times \langle \mathcal{C}|{\rm tr}_{V\otimes \bar{V}}\left[(\sigma_2\otimes \sigma_2)(L_1(u)\otimes L_1(-u))\cdots(L_{\frac{L}{2}}(u)\otimes L_{\frac{L}{2}}(-u))R(2u)R(-2u){\bf P}\right]\period\nonumber
\end{align}
Comparing \eqref{eq:rewriting2} and \eqref{eq:rewriting3}, we see that the right hand sides are symmetric with respect to $u\leftrightarrow -u$, which must also be true for the left hand sides. This leads to the equality we wanted to prove
\beq
\langle \mathcal{C}|T(u)=\langle \mathcal{C}|T(-u)\period
\eeq
\section{Conclusion\label{sec:conclusion}}
In this paper, we studied the crosscap states in integrable field theories and spin chains. The main outcome is the exact formula for the overlaps between the crosscap state and the energy eigenstates, \eqref{eq:fieldfinal} and \eqref{eq:spinnormalized}. As far as we know, our work is the first systematic analysis of the crosscap states in integrable systems and it opens a number of future directions, some of which we discuss below.
\vspace{-10pt}\paragraph{Generalizations.} On the integrable field theory side, an immediate next step is to generalize our formula to more general theories, such as theories with bound states and theories with non-diagonal scatterings. Such generalizations have been discussed for $g$-functions in \cite{Vu:2019qxt,Jiang:2019xdz,Jiang:2019zig} and we do not foresee fundamental obstacles in performing the same analysis for crosscaps. In particular, it would be interesting to study the WZNW model (cf.~\cite{Vu:2019qxt}) and test it against the CFT predictions at UV and IR fixed points.

On the spin-chain side, it would be interesting to find a generalization to higher-rank spin chains and derive determinant formulas for the overlaps. In higher-rank spin chains, one can consider more general crosscap states in which one combines the antipodal identification with the $\mathbb{Z}_2$ symmetry of the theory. In the case of $g$-functions, the detailed form of the determinants depends on the choice of the outer automorphism of the symmetry algebra \cite{deLeeuw:2015hxa,Jiang:2019xdz,Gombor:2020auk,Kristjansen:2020vbe}. It is natural to expect that this freedom also exists for the crosscaps and is related to the choice of the $\mathbb{Z}_2$ symmetry mentioned above. It would be interesting to work this out in examples and classify integrable crosscap states. Also, such generalizations would provide an ideal setup for studying the transformation property of the overlap formula under the change of gradings \cite{Kristjansen:2020vbe,Kristjansen:2021xno}, thanks to the absence of non-universal prefactors. 
\vspace{-12pt}\paragraph{Proof of the spin-chain overlap formula.} At a more technical level, it would be desirable to analytically prove our overlap formula \eqref{eq:overlapspin} for the spin-chain crosscap states. There are several possible strategies; mapping the system to a lattice model \cite{Foda:2015nfk,tsuchiya1998determinant}, using the algebraic Bethe ansatz and separation of variables \cite{Gombor:2021uxz}, directly working with the coordinate Bethe ansatz \cite{Jiang:2020sdw}, or developing a systematic algebraic approach as was recently achieved in \cite{Gombor:2021hmj} for the boundary states.
\vspace{-12pt}\paragraph{Staircase, $\mathbb{Z}_2$-orbifold and fermionization.} The $\mathbb{Z}_2$-orbifolded  staircase model, which we conjectured to describe the flows between $D$-series minimal models, deserves further study. For instance, it would be interesting to compute boundary entropies and  confirm the agreement with CFT. The same comment applies also to the fermionic staircase model, which we expect to describe the flows between fermionic minimal models. Note that the crosscap states in fermionic minimal models have not been fully analyzed in the literature, and it is by itself an interesting problem. More generally, the $\mathbb{Z}_2$-orbifold and the fermionization discussed in section \ref{sec:flow} can be applied to any integrable quantum field theory with $\mathbb{Z}_2$-symmetry. In particular, it would be interesting to study the fermionic version of the sinh-Gordon model, for which a direct Lagrangian description does not exist. (This is in contrast to the sine-Gordon model, whose fermionic version is the massive Thirring model). Another interesting question is to study generalizations such as $\mathbb{Z}_k$-orbifolding and $\mathbb{Z}_k$-parafermionization of integrable field theories. See \cite{Yao:2020dqx} for $\mathbb{Z}_k$-parafermionization of minimal models and \cite{Gaiotto:2020iye} for the groupoid structures of orbifolding and fermionization, and the connection to $3d$ TQFT.
\vspace{-12pt}\paragraph{$\boldsymbol{p}$-theorem?} We found that the crosscap entropy decreases monotonically along the RG flow of the staircase model, which contains infinitely many flows between $A$-series minimal models as its limit. On the other hand, the result for the $\mathbb{Z}_2$-orbifolded staircase model starts to increase in the deep infrared in the vicinity of the $\mathbb{Z}_2$-symmetry broken phase, where the theory becomes fully massive. However, we would like to emphasize that even for the orbifolded model the $p$-function decreases for most part of the flow, and in particular it is monotonic along the massless flows between $D$-series minimal models.
So far, the connection between the monotonicity and the spontaneous symmetry breaking is just an empirical observation and it is desirable to study other theories to understand it better. Even more interesting would be to prove the monotonicity of the crosscap entropy under certain assumptions\footnote{Another unconventional feature of the $p$-function is that it is not constant on the conformal manifold as can be seen explicitly in the compactified free boson CFT \cite{Tang:2018mgw}. We thank also Yifan Wang for bringing this to our attention.} by generalizing the proofs for the $g$-function \cite{Friedan:2003yc,Casini:2016fgb,Cuomo:2021rkm}.
\vspace{-12pt}\paragraph{Alternative formulation.} Recently, there have been several works which attempted to derive an alternative representation for the $g$-function; either based on the integral equation of the TBA type (called the ``Tracy-Widom TBA'') \cite{Caetano:2020dyp} or based on separation of variables \cite{Caetano:2020dyp,Cavaglia:2021mft}. One unsatisfactory feature of all these works is that they only give a ratio of Fredholm determinants and one has to multiply a non-universal prefactor in order to get the full $g$-function. As we saw in this paper, such a non-universal prefactor is absent in the crosscap overlap. This makes the crosscap overlap an ideal target for these new approaches and motivates further studies.
\vspace{-12pt}\paragraph{Application to string theory and holography.} Historically, one of the motivations for studying the crosscap states in 2d CFT came from string theory on orientifold backgrounds. Our results generalize the construction of crosscaps in 2d CFT to integrable theories, and are potentially useful for string theory for which the worldsheet dynamics is integrable, e.g.~type IIB string theory on AdS$_5\times$S$^5$. Famous examples of orientifolds of AdS$_5\times$S$^5$ are those dual to $\mathcal{N}=4$ SYM with $SO(N)$ or $Sp(N)$ gauge groups \cite{Witten:1998xy} (see also the discussion in the context of integrability of $\mathcal{N}=4$ SYM \cite{Caputa:2010ep}). However, the crosscap overlaps vanish in these examples since they are dual to vacuum one-point functions of local operators, which are zero due to conformal symmetry. In order to apply our techniques and obtain nontrivial results, one needs to find a setup in which the conformal symmetry is partially broken by the orientifold. Work in this direction is in progress \cite{inprogress}. Finding such a setup in AdS/CFT is important also from a broader point of view, as this could provide a non-perturbative definition of orientifold planes, which play a key role in various model buildings in string phenomenology including the construction of de Sitter vacua, see e.g.~\cite{Cordova:2018dbb,Cordova:2019cvf}.
\vspace{-12pt}\paragraph{Conclusion.}
Needless to say, the list above is not exhaustive, and there are certainly many other open questions. For instance, the crosscap states are important for understanding topological phases in $2+1$ dimensions protected by the reflection symmetry (see e.g.~\cite{Cho:2015ega}). It would be interesting see if our construction in spin chains has any implication on it. We hope that our findings will pave the way for exciting and unexpected future developments.
\subsubsection*{Acknowledgement}
We thank Leonardo Rastelli for discussions on related topics. We also thank Ivan Kostov, Shu-Heng Shao and especially Yifan Wang for valuable comments on the manuscript. 
\bibliographystyle{JHEP}
\bibliography{crosscap}

\providecommand{\href}[2]{#2}\begingroup\raggedright\begin{thebibliography}{100}

\bibitem{kondo1964resistance}
J.~Kondo, \emph{Resistance minimum in dilute magnetic alloys}, {\emph{Progress
  of theoretical physics} {\bfseries 32} (1964) 37}.

\bibitem{Ishibashi:1988kg}
N.~Ishibashi, \emph{{The Boundary and Crosscap States in Conformal Field
  Theories}}, \href{https://doi.org/10.1142/S0217732389000320}{\emph{Mod. Phys.
  Lett. A} {\bfseries 4} (1989) 251}.

\bibitem{Cardy:1989ir}
J.~L. Cardy, \emph{{Boundary Conditions, Fusion Rules and the Verlinde
  Formula}}, \href{https://doi.org/10.1016/0550-3213(89)90521-X}{\emph{Nucl.
  Phys. B} {\bfseries 324} (1989) 581}.

\bibitem{Cardy:1991tv}
J.~L. Cardy and D.~C. Lewellen, \emph{{Bulk and boundary operators in conformal
  field theory}},
  \href{https://doi.org/10.1016/0370-2693(91)90828-E}{\emph{Phys. Lett. B}
  {\bfseries 259} (1991) 274}.

\bibitem{Behrend:1998fd}
R.~E. Behrend, P.~A. Pearce, V.~B. Petkova and J.-B. Zuber, \emph{{On the
  classification of bulk and boundary conformal field theories}},
  \href{https://doi.org/10.1016/S0370-2693(98)01374-4}{\emph{Phys. Lett. B}
  {\bfseries 444} (1998) 163}
  [\href{https://arxiv.org/abs/hep-th/9809097}{{\ttfamily hep-th/9809097}}].

\bibitem{Behrend:1999bn}
R.~E. Behrend, P.~A. Pearce, V.~B. Petkova and J.-B. Zuber, \emph{{Boundary
  conditions in rational conformal field theories}},
  \href{https://doi.org/10.1016/S0550-3213(99)00592-1}{\emph{Nucl. Phys. B}
  {\bfseries 570} (2000) 525}
  [\href{https://arxiv.org/abs/hep-th/9908036}{{\ttfamily hep-th/9908036}}].

\bibitem{Ghoshal:1993tm}
S.~Ghoshal and A.~B. Zamolodchikov, \emph{{Boundary S matrix and boundary state
  in two-dimensional integrable quantum field theory}},
  \href{https://doi.org/10.1142/S0217751X94001552}{\emph{Int. J. Mod. Phys.}
  {\bfseries A9} (1994) 3841}
  [\href{https://arxiv.org/abs/hep-th/9306002}{{\ttfamily hep-th/9306002}}].

\bibitem{Dorey:1999cj}
P.~Dorey, I.~Runkel, R.~Tateo and G.~Watts, \emph{{g function flow in perturbed
  boundary conformal field theories}},
  \href{https://doi.org/10.1016/S0550-3213(99)00772-5}{\emph{Nucl. Phys.}
  {\bfseries B578} (2000) 85}
  [\href{https://arxiv.org/abs/hep-th/9909216}{{\ttfamily hep-th/9909216}}].

\bibitem{Dorey:2009vg}
P.~Dorey, C.~Rim and R.~Tateo, \emph{{Exact g-function flow between conformal
  field theories}},
  \href{https://doi.org/10.1016/j.nuclphysb.2010.03.010}{\emph{Nucl. Phys.}
  {\bfseries B834} (2010) 485}
  [\href{https://arxiv.org/abs/0911.4969}{{\ttfamily 0911.4969}}].

\bibitem{Dorey:2004xk}
P.~Dorey, D.~Fioravanti, C.~Rim and R.~Tateo, \emph{{Integrable quantum field
  theory with boundaries: The Exact g function}},
  \href{https://doi.org/10.1016/j.nuclphysb.2004.06.045}{\emph{Nucl. Phys.}
  {\bfseries B696} (2004) 445}
  [\href{https://arxiv.org/abs/hep-th/0404014}{{\ttfamily hep-th/0404014}}].

\bibitem{Dorey:2010ub}
P.~Dorey, R.~Tateo and R.~Wilbourne, \emph{{Exact g-function flows from the
  staircase model}},
  \href{https://doi.org/10.1016/j.nuclphysb.2010.10.009}{\emph{Nucl. Phys. B}
  {\bfseries 843} (2011) 724}
  [\href{https://arxiv.org/abs/1008.1190}{{\ttfamily 1008.1190}}].

\bibitem{Affleck:1990by}
I.~Affleck and A.~W.~W. Ludwig, \emph{{The Kondo effect, conformal field theory
  and fusion rules}},
  \href{https://doi.org/10.1016/0550-3213(91)90109-B}{\emph{Nucl. Phys. B}
  {\bfseries 352} (1991) 849}.

\bibitem{Affleck:1990iv}
I.~Affleck and A.~W.~W. Ludwig, \emph{{Critical theory of overscreened Kondo
  fixed points}},
  \href{https://doi.org/10.1016/0550-3213(91)90419-X}{\emph{Nucl. Phys. B}
  {\bfseries 360} (1991) 641}.

\bibitem{Andrei:1980fv}
N.~Andrei, \emph{{Diagonalization of the Kondo Hamiltonian}},
  \href{https://doi.org/10.1103/PhysRevLett.45.379}{\emph{Phys. Rev. Lett.}
  {\bfseries 45} (1980) 379}.

\bibitem{wiegmann1981exact}
P.~Wiegmann, \emph{Exact solution of the sd exchange model (kondo problem)},
  {\emph{Journal of Physics C: Solid State Physics} {\bfseries 14} (1981)
  1463}.

\bibitem{Calabrese:2006rx}
P.~Calabrese and J.~L. Cardy, \emph{{Time-dependence of correlation functions
  following a quantum quench}},
  \href{https://doi.org/10.1103/PhysRevLett.96.136801}{\emph{Phys. Rev. Lett.}
  {\bfseries 96} (2006) 136801}
  [\href{https://arxiv.org/abs/cond-mat/0601225}{{\ttfamily
  cond-mat/0601225}}].

\bibitem{Calabrese:2007mtj}
P.~Calabrese and J.~Cardy, \emph{{Entanglement and correlation functions
  following a local quench: a conformal field theory approach}},
  \href{https://doi.org/10.1088/1742-5468/2007/10/P10004}{\emph{J. Stat. Mech.}
  {\bfseries 0710} (2007) P10004}
  [\href{https://arxiv.org/abs/0708.3750}{{\ttfamily 0708.3750}}].

\bibitem{Caux:2013ra}
J.-S. Caux and F.~H.~L. Essler, \emph{{Time evolution of local observables
  after quenching to an integrable model}},
  \href{https://doi.org/10.1103/PhysRevLett.110.257203}{\emph{Phys. Rev. Lett.}
  {\bfseries 110} (2013) 257203}
  [\href{https://arxiv.org/abs/1301.3806}{{\ttfamily 1301.3806}}].

\bibitem{Caux:2016esd}
J.-S. Caux, \emph{{The Quench Action}},
  \href{https://doi.org/10.1088/1742-5468/2016/06/064006}{\emph{J. Stat. Mech.}
  {\bfseries 1606} (2016) 064006}
  [\href{https://arxiv.org/abs/1603.04689}{{\ttfamily 1603.04689}}].

\bibitem{brockmann2014gaudin}
M.~Brockmann, J.~De~Nardis, B.~Wouters and J.-S. Caux, \emph{A gaudin-like
  determinant for overlaps of n{\'e}el and xxz bethe states}, {\emph{Journal of
  Physics A: Mathematical and Theoretical} {\bfseries 47} (2014) 145003}.

\bibitem{brockmann2014neel}
M.~Brockmann, J.~De~Nardis, B.~Wouters and J.-S. Caux, \emph{N{\'e}el-xxz state
  overlaps: odd particle numbers and lieb--liniger scaling limit},
  {\emph{Journal of Physics A: Mathematical and Theoretical} {\bfseries 47}
  (2014) 345003}.

\bibitem{pozsgay2014overlaps}
B.~Pozsgay, \emph{Overlaps between eigenstates of the xxz spin-1/2 chain and a
  class of simple product states}, {\emph{Journal of Statistical Mechanics:
  Theory and Experiment} {\bfseries 2014} (2014) P06011}.

\bibitem{Foda:2015nfk}
O.~Foda and K.~Zarembo, \emph{{Overlaps of partial N\'eel states and Bethe
  states}}, \href{https://doi.org/10.1088/1742-5468/2016/02/023107}{\emph{J.
  Stat. Mech.} {\bfseries 1602} (2016) 023107}
  [\href{https://arxiv.org/abs/1512.02533}{{\ttfamily 1512.02533}}].

\bibitem{Pozsgay:2018ixm}
B.~Pozsgay, \emph{{Overlaps with arbitrary two-site states in the XXZ spin
  chain}}, \href{https://doi.org/10.1088/1742-5468/aabbe1}{\emph{J. Stat.
  Mech.} {\bfseries 1805} (2018) 053103}
  [\href{https://arxiv.org/abs/1801.03838}{{\ttfamily 1801.03838}}].

\bibitem{Piroli:2018ksf}
L.~Piroli, E.~Vernier, P.~Calabrese and B.~Pozsgay, \emph{{Integrable quenches
  in nested spin chains I: the exact steady states}},
  \href{https://doi.org/10.1088/1742-5468/ab1c51}{\emph{J. Stat. Mech.}
  {\bfseries 1906} (2019) 063103}
  [\href{https://arxiv.org/abs/1811.00432}{{\ttfamily 1811.00432}}].

\bibitem{Pozsgay:2018dzs}
B.~Pozsgay, L.~Piroli and E.~Vernier, \emph{{Integrable Matrix Product States
  from boundary integrability}},
  \href{https://doi.org/10.21468/SciPostPhys.6.5.062}{\emph{SciPost Phys.}
  {\bfseries 6} (2019) 062} [\href{https://arxiv.org/abs/1812.11094}{{\ttfamily
  1812.11094}}].

\bibitem{DeLeeuw:2019ohp}
M.~De~Leeuw, T.~Gombor, C.~Kristjansen, G.~Linardopoulos and B.~Pozsgay,
  \emph{{Spin Chain Overlaps and the Twisted Yangian}},
  \href{https://doi.org/10.1007/JHEP01(2020)176}{\emph{JHEP} {\bfseries 01}
  (2020) 176} [\href{https://arxiv.org/abs/1912.09338}{{\ttfamily
  1912.09338}}].

\bibitem{Piroli:2018amn}
L.~Piroli, B.~Pozsgay and E.~Vernier, \emph{{Non-analytic behavior of the
  Loschmidt echo in XXZ spin chains: Exact results}},
  \href{https://doi.org/10.1016/j.nuclphysb.2018.06.015}{\emph{Nucl. Phys. B}
  {\bfseries 933} (2018) 454}
  [\href{https://arxiv.org/abs/1803.04380}{{\ttfamily 1803.04380}}].

\bibitem{Maldacena:1997re}
J.~M. Maldacena, \emph{{The Large N limit of superconformal field theories and
  supergravity}}, \href{https://doi.org/10.1023/A:1026654312961}{\emph{Adv.
  Theor. Math. Phys.} {\bfseries 2} (1998) 231}
  [\href{https://arxiv.org/abs/hep-th/9711200}{{\ttfamily hep-th/9711200}}].

\bibitem{deLeeuw:2015hxa}
M.~de~Leeuw, C.~Kristjansen and K.~Zarembo, \emph{{One-point Functions in
  Defect CFT and Integrability}},
  \href{https://doi.org/10.1007/JHEP08(2015)098}{\emph{JHEP} {\bfseries 08}
  (2015) 098} [\href{https://arxiv.org/abs/1506.06958}{{\ttfamily
  1506.06958}}].

\bibitem{Buhl-Mortensen:2015gfd}
I.~Buhl-Mortensen, M.~de~Leeuw, C.~Kristjansen and K.~Zarembo, \emph{{One-point
  Functions in AdS/dCFT from Matrix Product States}},
  \href{https://doi.org/10.1007/JHEP02(2016)052}{\emph{JHEP} {\bfseries 02}
  (2016) 052} [\href{https://arxiv.org/abs/1512.02532}{{\ttfamily
  1512.02532}}].

\bibitem{Buhl-Mortensen:2017ind}
I.~Buhl-Mortensen, M.~de~Leeuw, A.~C. Ipsen, C.~Kristjansen and M.~Wilhelm,
  \emph{{Asymptotic One-Point Functions in Gauge-String Duality with Defects}},
  \href{https://doi.org/10.1103/PhysRevLett.119.261604}{\emph{Phys. Rev. Lett.}
  {\bfseries 119} (2017) 261604}
  [\href{https://arxiv.org/abs/1704.07386}{{\ttfamily 1704.07386}}].

\bibitem{Komatsu:2020sup}
S.~Komatsu and Y.~Wang, \emph{{Non-perturbative defect one-point functions in
  planar $\mathcal{N}=4$ super-Yang-Mills}},
  \href{https://doi.org/10.1016/j.nuclphysb.2020.115120}{\emph{Nucl. Phys. B}
  {\bfseries 958} (2020) 115120}
  [\href{https://arxiv.org/abs/2004.09514}{{\ttfamily 2004.09514}}].

\bibitem{Gombor:2020kgu}
T.~Gombor and Z.~Bajnok, \emph{{Boundary states, overlaps, nesting and
  bootstrapping AdS/dCFT}},
  \href{https://doi.org/10.1007/JHEP10(2020)123}{\emph{JHEP} {\bfseries 10}
  (2020) 123} [\href{https://arxiv.org/abs/2004.11329}{{\ttfamily
  2004.11329}}].

\bibitem{Gombor:2020auk}
T.~Gombor and Z.~Bajnok, \emph{{Boundary state bootstrap and asymptotic
  overlaps in AdS/dCFT}},
  \href{https://doi.org/10.1007/JHEP03(2021)222}{\emph{JHEP} {\bfseries 03}
  (2021) 222} [\href{https://arxiv.org/abs/2006.16151}{{\ttfamily
  2006.16151}}].

\bibitem{Jiang:2019xdz}
Y.~Jiang, S.~Komatsu and E.~Vescovi, \emph{{Structure constants in $
  \mathcal{N} $ = 4 SYM at finite coupling as worldsheet g-function}},
  \href{https://doi.org/10.1007/JHEP07(2020)037}{\emph{JHEP} {\bfseries 07}
  (2020) 037} [\href{https://arxiv.org/abs/1906.07733}{{\ttfamily
  1906.07733}}].

\bibitem{Jiang:2019zig}
Y.~Jiang, S.~Komatsu and E.~Vescovi, \emph{{Exact Three-Point Functions of
  Determinant Operators in Planar $N=4$ Supersymmetric Yang-Mills Theory}},
  \href{https://doi.org/10.1103/PhysRevLett.123.191601}{\emph{Phys. Rev. Lett.}
  {\bfseries 123} (2019) 191601}
  [\href{https://arxiv.org/abs/1907.11242}{{\ttfamily 1907.11242}}].

\bibitem{Yang:2021hrl}
P.~Yang, Y.~Jiang, S.~Komatsu and J.-B. Wu, \emph{{Three-Point Functions in
  ABJM and Bethe Ansatz}},  \href{https://arxiv.org/abs/2103.15840}{{\ttfamily
  2103.15840}}.

\bibitem{Sagnotti:1987tw}
A.~Sagnotti, \emph{{Open Strings and their Symmetry Groups}},  in \emph{{NATO
  Advanced Summer Institute on Nonperturbative Quantum Field Theory (Cargese
  Summer Institute)}}, 9, 1987,
  \href{https://arxiv.org/abs/hep-th/0208020}{{\ttfamily hep-th/0208020}}.

\bibitem{Fioravanti:1993hf}
D.~Fioravanti, G.~Pradisi and A.~Sagnotti, \emph{{Sewing constraints and
  nonorientable open strings}},
  \href{https://doi.org/10.1016/0370-2693(94)90255-0}{\emph{Phys. Lett. B}
  {\bfseries 321} (1994) 349}
  [\href{https://arxiv.org/abs/hep-th/9311183}{{\ttfamily hep-th/9311183}}].

\bibitem{Pradisi:1996yd}
G.~Pradisi, A.~Sagnotti and Y.~S. Stanev, \emph{{Completeness conditions for
  boundary operators in 2-D conformal field theory}},
  \href{https://doi.org/10.1016/0370-2693(96)00578-3}{\emph{Phys. Lett. B}
  {\bfseries 381} (1996) 97}
  [\href{https://arxiv.org/abs/hep-th/9603097}{{\ttfamily hep-th/9603097}}].

\bibitem{Angelantonj:2002ct}
C.~Angelantonj and A.~Sagnotti, \emph{{Open strings}},
  \href{https://doi.org/10.1016/S0370-1573(02)00273-9}{\emph{Phys. Rept.}
  {\bfseries 371} (2002) 1}
  [\href{https://arxiv.org/abs/hep-th/0204089}{{\ttfamily hep-th/0204089}}].

\bibitem{Affleck:1991tk}
I.~Affleck and A.~W.~W. Ludwig, \emph{{Universal noninteger 'ground state
  degeneracy' in critical quantum systems}},
  \href{https://doi.org/10.1103/PhysRevLett.67.161}{\emph{Phys. Rev. Lett.}
  {\bfseries 67} (1991) 161}.

\bibitem{Tu:2017wks}
H.-H. Tu, \emph{{Universal entropy of conformal critical theories on a Klein
  bottle}}, \href{https://doi.org/10.1103/PhysRevLett.119.261603}{\emph{Phys.
  Rev. Lett.} {\bfseries 119} (2017) 261603}
  [\href{https://arxiv.org/abs/1707.05812}{{\ttfamily 1707.05812}}].

\bibitem{Garcia-Compean:2018ury}
H.~Garc\'\i{}a-Compe\'an and N.~Quiroz, \emph{{On the topological and crosscap
  entropies in non-oriented RCFTs}},
  \href{https://doi.org/10.1140/epjp/s13360-021-01878-y}{\emph{Eur. Phys. J.
  Plus} {\bfseries 136} (2021) 881}
  [\href{https://arxiv.org/abs/1811.07238}{{\ttfamily 1811.07238}}].

\bibitem{Tang:2018mgw}
W.~Tang, X.~C. Xie, L.~Wang and H.-H. Tu, \emph{{Klein bottle entropy of
  compactified boson conformal field theory}},
  \href{https://doi.org/10.1103/PhysRevB.99.115105}{\emph{Phys. Rev. B}
  {\bfseries 99} (2019) 115105}
  [\href{https://arxiv.org/abs/1805.01300}{{\ttfamily 1805.01300}}].

\bibitem{Cuomo:2021rkm}
G.~Cuomo, Z.~Komargodski and A.~Raviv-Moshe, \emph{{Renormalization Group Flows
  on Line Defects}},  \href{https://arxiv.org/abs/2108.01117}{{\ttfamily
  2108.01117}}.

\bibitem{LeClair:1995uf}
A.~LeClair, G.~Mussardo, H.~Saleur and S.~Skorik, \emph{{Boundary energy and
  boundary states in integrable quantum field theories}},
  \href{https://doi.org/10.1016/0550-3213(95)00435-U}{\emph{Nucl. Phys.}
  {\bfseries B453} (1995) 581}
  [\href{https://arxiv.org/abs/hep-th/9503227}{{\ttfamily hep-th/9503227}}].

\bibitem{Woynarovich:2004gc}
F.~Woynarovich, \emph{{O(1) contribution of saddle point fluctuations to the
  free energy of Bethe Ansatz systems}},
  \href{https://doi.org/10.1016/j.nuclphysb.2004.08.043}{\emph{Nucl. Phys.}
  {\bfseries B700} (2004) 331}
  [\href{https://arxiv.org/abs/cond-mat/0402129}{{\ttfamily
  cond-mat/0402129}}].

\bibitem{Pozsgay:2010tv}
B.~Pozsgay, \emph{{On O(1) contributions to the free energy in Bethe Ansatz
  systems: The Exact g-function}},
  \href{https://doi.org/10.1007/JHEP08(2010)090}{\emph{JHEP} {\bfseries 08}
  (2010) 090} [\href{https://arxiv.org/abs/1003.5542}{{\ttfamily 1003.5542}}].

\bibitem{Zamolodchikov:1989cf}
A.~B. Zamolodchikov, \emph{{Thermodynamic Bethe Ansatz in Relativistic Models.
  Scaling Three State Potts and Lee-yang Models}},
  \href{https://doi.org/10.1016/0550-3213(90)90333-9}{\emph{Nucl. Phys. B}
  {\bfseries 342} (1990) 695}.

\bibitem{Kostov:2018dmi}
I.~Kostov, D.~Serban and D.-L. Vu, \emph{{Boundary TBA, trees and loops}},
  \href{https://doi.org/10.1016/j.nuclphysb.2019.114817}{\emph{Nucl. Phys. B}
  {\bfseries 949} (2019) 114817}
  [\href{https://arxiv.org/abs/1809.05705}{{\ttfamily 1809.05705}}].

\bibitem{Kostov:2019sgu}
I.~Kostov, \emph{{Effective Quantum Field Theory for the Thermodynamical Bethe
  Ansatz}}, \href{https://doi.org/10.1007/JHEP02(2020)043}{\emph{JHEP}
  {\bfseries 02} (2020) 043}
  [\href{https://arxiv.org/abs/1911.07343}{{\ttfamily 1911.07343}}].

\bibitem{Dorey:1996re}
P.~Dorey and R.~Tateo, \emph{{Excited states by analytic continuation of TBA
  equations}}, \href{https://doi.org/10.1016/S0550-3213(96)00516-0}{\emph{Nucl.
  Phys.} {\bfseries B482} (1996) 639}
  [\href{https://arxiv.org/abs/hep-th/9607167}{{\ttfamily hep-th/9607167}}].

\bibitem{Kristjansen:2010kg}
C.~Kristjansen, \emph{{Review of AdS/CFT Integrability, Chapter IV.1: Aspects
  of Non-Planarity}},
  \href{https://doi.org/10.1007/s11005-011-0514-9}{\emph{Lett. Math. Phys.}
  {\bfseries 99} (2012) 349} [\href{https://arxiv.org/abs/1012.3997}{{\ttfamily
  1012.3997}}].

\bibitem{Piroli:2017sei}
L.~Piroli, B.~Pozsgay and E.~Vernier, \emph{{What is an integrable quench?}},
  \href{https://doi.org/10.1016/j.nuclphysb.2017.10.012}{\emph{Nucl. Phys. B}
  {\bfseries 925} (2017) 362}
  [\href{https://arxiv.org/abs/1709.04796}{{\ttfamily 1709.04796}}].

\bibitem{Zamolodchikov:2006vf}
A.~B. Zamolodchikov, \emph{Resonance factorized scattering and roaming
  trajectories}, .

\bibitem{Dorey:1996gd}
P.~Dorey, \emph{{Exact S matrices}},  in \emph{{Eotvos Summer School in
  Physics: Conformal Field Theories and Integrable Models}}, pp.~85--125, 8,
  1996, \href{https://arxiv.org/abs/hep-th/9810026}{{\ttfamily
  hep-th/9810026}}.

\bibitem{Tsiares:2020ewp}
I.~Tsiares, \emph{{Universal Dynamics in Non-Orientable CFT$_2$}},
  \href{https://arxiv.org/abs/2011.09250}{{\ttfamily 2011.09250}}.

\bibitem{Blumenhagen:2009zz}
R.~Blumenhagen and E.~Plauschinn, \emph{{Introduction to conformal field
  theory}: {with applications to String theory}}, vol.~779. 2009,
  \href{https://doi.org/10.1007/978-3-642-00450-6}{10.1007/978-3-642-00450-6}.

\bibitem{OnogiIshibashi}
T.~Onogi and N.~Ishibashi, \emph{Conformal field theories on surfaces with
  boundaries and crosscaps},
  \href{https://doi.org/10.1142/S0217732389000228}{\emph{Modern Physics Letters
  A} {\bfseries 04} (1989) 161}
  [\href{https://arxiv.org/abs/https://doi.org/10.1142/S0217732389000228}{{\ttfamily
  https://doi.org/10.1142/S0217732389000228}}].

\bibitem{DiFrancesco:1997nk}
P.~Di~Francesco, P.~Mathieu and D.~Senechal, \emph{{Conformal Field Theory}},
  Graduate Texts in Contemporary Physics. Springer-Verlag, New York, 1997,
  \href{https://doi.org/10.1007/978-1-4612-2256-9}{10.1007/978-1-4612-2256-9}.

\bibitem{Hsieh:2020uwb}
C.-T. Hsieh, Y.~Nakayama and Y.~Tachikawa, \emph{{Fermionic minimal models}},
  \href{https://doi.org/10.1103/PhysRevLett.126.195701}{\emph{Phys. Rev. Lett.}
  {\bfseries 126} (2021) 195701}
  [\href{https://arxiv.org/abs/2002.12283}{{\ttfamily 2002.12283}}].

\bibitem{Klassen:1991dz}
T.~R. Klassen and E.~Melzer, \emph{{RG flows in the D series of minimal CFTs}},
  \href{https://doi.org/10.1016/0550-3213(93)90415-L}{\emph{Nucl. Phys. B}
  {\bfseries 400} (1993) 547}
  [\href{https://arxiv.org/abs/hep-th/9110047}{{\ttfamily hep-th/9110047}}].

\bibitem{Vafa:1989ih}
C.~Vafa, \emph{{Quantum Symmetries of String Vacua}},
  \href{https://doi.org/10.1142/S0217732389001842}{\emph{Mod. Phys. Lett. A}
  {\bfseries 4} (1989) 1615}.

\bibitem{Chai:2020onq}
N.~Chai, S.~Chaudhuri, C.~Choi, Z.~Komargodski, E.~Rabinovici and M.~Smolkin,
  \emph{{Symmetry Breaking at All Temperatures}},
  \href{https://doi.org/10.1103/PhysRevLett.125.131603}{\emph{Phys. Rev. Lett.}
  {\bfseries 125} (2020) 131603}.

\bibitem{Kitaev:2000nmw}
A.~Y. Kitaev, \emph{{Unpaired Majorana fermions in quantum wires}},
  \href{https://doi.org/10.1070/1063-7869/44/10S/S29}{\emph{Phys. Usp.}
  {\bfseries 44} (2001) 131}
  [\href{https://arxiv.org/abs/cond-mat/0010440}{{\ttfamily
  cond-mat/0010440}}].

\bibitem{Petkova:1988cy}
V.~B. Petkova, \emph{{Two-dimensional (Half) Integer Spin Conformal Theories
  With Central Charge $C < 1$}},
  \href{https://doi.org/10.1142/S0217751X88001235}{\emph{Int. J. Mod. Phys. A}
  {\bfseries 3} (1988) 2945}.

\bibitem{Furlan:1989ra}
P.~Furlan, A.~C. Ganchev and V.~B. Petkova, \emph{{Fusion Matrices and $C < 1$
  (Quasi)local Conformal Theories}},
  \href{https://doi.org/10.1142/S0217751X90001252}{\emph{Int. J. Mod. Phys. A}
  {\bfseries 5} (1990) 2721}.

\bibitem{Karch:2019lnn}
A.~Karch, D.~Tong and C.~Turner, \emph{{A Web of 2d Dualities: ${\bf Z}_2$
  Gauge Fields and Arf Invariants}},
  \href{https://doi.org/10.21468/SciPostPhys.7.1.007}{\emph{SciPost Phys.}
  {\bfseries 7} (2019) 007} [\href{https://arxiv.org/abs/1902.05550}{{\ttfamily
  1902.05550}}].

\bibitem{Runkel:2020zgg}
I.~Runkel and G.~M.~T. Watts, \emph{{Fermionic CFTs and classifying algebras}},
  \href{https://doi.org/10.1007/JHEP06(2020)025}{\emph{JHEP} {\bfseries 06}
  (2020) 025} [\href{https://arxiv.org/abs/2001.05055}{{\ttfamily
  2001.05055}}].

\bibitem{Kulp:2020iet}
J.~Kulp, \emph{{Two More Fermionic Minimal Models}},
  \href{https://doi.org/10.1007/JHEP03(2021)124}{\emph{JHEP} {\bfseries 03}
  (2021) 124} [\href{https://arxiv.org/abs/2003.04278}{{\ttfamily
  2003.04278}}].

\bibitem{Smith:2021luc}
P.~B. Smith, \emph{{Boundary States and Anomalous Symmetries of Fermionic
  Minimal Models}},  \href{https://arxiv.org/abs/2102.02203}{{\ttfamily
  2102.02203}}.

\bibitem{Fukusumi:2021zme}
Y.~Fukusumi, Y.~Tachikawa and Y.~Zheng, \emph{{Fermionization and boundary
  states in 1+1 dimensions}},
  \href{https://doi.org/10.21468/SciPostPhys.11.4.082}{\emph{SciPost Phys.}
  {\bfseries 11} (2021) 082}
  [\href{https://arxiv.org/abs/2103.00746}{{\ttfamily 2103.00746}}].

\bibitem{Ebisu:2021acm}
H.~Ebisu and M.~Watanabe, \emph{{Fermionization of conformal boundary states}},
   \href{https://arxiv.org/abs/2103.01101}{{\ttfamily 2103.01101}}.

\bibitem{Klassen:1992eq}
T.~R. Klassen and E.~Melzer, \emph{{Sine-Gordon not equal to massive Thirring,
  and related heresies}},
  \href{https://doi.org/10.1142/S0217751X93001703}{\emph{Int. J. Mod. Phys. A}
  {\bfseries 8} (1993) 4131}
  [\href{https://arxiv.org/abs/hep-th/9206114}{{\ttfamily hep-th/9206114}}].

\bibitem{Gliozzi:1976jf}
F.~Gliozzi, J.~Scherk and D.~I. Olive, \emph{{Supergravity and the Spinor Dual
  Model}}, \href{https://doi.org/10.1016/0370-2693(76)90183-0}{\emph{Phys.
  Lett. B} {\bfseries 65} (1976) 282}.

\bibitem{Gliozzi:1976qd}
F.~Gliozzi, J.~Scherk and D.~I. Olive, \emph{{Supersymmetry, Supergravity
  Theories and the Dual Spinor Model}},
  \href{https://doi.org/10.1016/0550-3213(77)90206-1}{\emph{Nucl. Phys. B}
  {\bfseries 122} (1977) 253}.

\bibitem{Seiberg:1986by}
N.~Seiberg and E.~Witten, \emph{{Spin Structures in String Theory}},
  \href{https://doi.org/10.1016/0550-3213(86)90297-X}{\emph{Nucl. Phys. B}
  {\bfseries 276} (1986) 272}.

\bibitem{Skrzypek:2021eue}
T.~Skrzypek and A.~A. Tseytlin, \emph{{On type 0 string theory in solvable RR
  backgrounds}},  \href{https://arxiv.org/abs/2110.14683}{{\ttfamily
  2110.14683}}.

\bibitem{Kaidi:2019pzj}
J.~Kaidi, J.~Parra-Martinez and Y.~Tachikawa, \emph{{Classification of String
  Theories via Topological Phases}},
  \href{https://doi.org/10.1103/PhysRevLett.124.121601}{\emph{Phys. Rev. Lett.}
  {\bfseries 124} (2020) 121601}
  [\href{https://arxiv.org/abs/1908.04805}{{\ttfamily 1908.04805}}].

\bibitem{Kaidi:2019tyf}
J.~Kaidi, J.~Parra-Martinez, Y.~Tachikawa and w.~a. m. a. b.~A. Debray,
  \emph{{Topological Superconductors on Superstring Worldsheets}},
  \href{https://doi.org/10.21468/SciPostPhys.9.1.010}{\emph{SciPost Phys.}
  {\bfseries 9} (2020) 10} [\href{https://arxiv.org/abs/1911.11780}{{\ttfamily
  1911.11780}}].

\bibitem{Mussardo:1999aj}
G.~Mussardo and P.~Simon, \emph{{Bosonic type S matrix, vacuum instability and
  CDD ambiguities}},
  \href{https://doi.org/10.1016/S0550-3213(99)00806-8}{\emph{Nucl. Phys. B}
  {\bfseries 578} (2000) 527}
  [\href{https://arxiv.org/abs/hep-th/9903072}{{\ttfamily hep-th/9903072}}].

\bibitem{Cordova:2021fnr}
L.~C\'ordova, S.~Negro and F.~I.~S. Massolo, \emph{{Thermodynamic Bethe Ansatz
  past turning points: the (elliptic) sinh-Gordon model}},
  \href{https://arxiv.org/abs/2110.14666}{{\ttfamily 2110.14666}}.

\bibitem{Faddeev:1996iy}
L.~D. Faddeev, \emph{{How algebraic Bethe ansatz works for integrable model}},
  in \emph{{Les Houches School of Physics: Astrophysical Sources of
  Gravitational Radiation}}, pp.~pp. 149--219, 5, 1996,
  \href{https://arxiv.org/abs/hep-th/9605187}{{\ttfamily hep-th/9605187}}.

\bibitem{Korepin:1993kvr}
V.~E. Korepin, N.~M. Bogoliubov and A.~G. Izergin, \emph{{Quantum Inverse
  Scattering Method and Correlation Functions}}, Cambridge Monographs on
  Mathematical Physics. Cambridge University Press, Cambridge, 1993,
  \href{https://doi.org/10.1017/CBO9780511628832}{10.1017/CBO9780511628832}.

\bibitem{Beisert:2004ry}
N.~Beisert, \emph{{The Dilatation operator of N=4 super Yang-Mills theory and
  integrability}},
  \href{https://doi.org/10.1016/j.physrep.2004.09.007}{\emph{Phys. Rept.}
  {\bfseries 405} (2004) 1}
  [\href{https://arxiv.org/abs/hep-th/0407277}{{\ttfamily hep-th/0407277}}].

\bibitem{Braun:1998id}
V.~M. Braun, S.~E. Derkachov and A.~N. Manashov, \emph{{Integrability of three
  particle evolution equations in QCD}},
  \href{https://doi.org/10.1103/PhysRevLett.81.2020}{\emph{Phys. Rev. Lett.}
  {\bfseries 81} (1998) 2020}
  [\href{https://arxiv.org/abs/hep-ph/9805225}{{\ttfamily hep-ph/9805225}}].

\bibitem{Braun:1999te}
V.~M. Braun, S.~E. Derkachov, G.~P. Korchemsky and A.~N. Manashov,
  \emph{{Baryon distribution amplitudes in QCD}},
  \href{https://doi.org/10.1016/S0550-3213(99)00265-5}{\emph{Nucl. Phys. B}
  {\bfseries 553} (1999) 355}
  [\href{https://arxiv.org/abs/hep-ph/9902375}{{\ttfamily hep-ph/9902375}}].

\bibitem{Belitsky:1999qh}
A.~V. Belitsky, \emph{{Fine structure of spectrum of twist - three operators in
  QCD}}, \href{https://doi.org/10.1016/S0370-2693(99)00326-3}{\emph{Phys. Lett.
  B} {\bfseries 453} (1999) 59}
  [\href{https://arxiv.org/abs/hep-ph/9902361}{{\ttfamily hep-ph/9902361}}].

\bibitem{Belitsky:1999ru}
A.~V. Belitsky, \emph{{Integrability and WKB solution of twist - three
  evolution equations}},
  \href{https://doi.org/10.1016/S0550-3213(99)00402-2}{\emph{Nucl. Phys. B}
  {\bfseries 558} (1999) 259}
  [\href{https://arxiv.org/abs/hep-ph/9903512}{{\ttfamily hep-ph/9903512}}].

\bibitem{Nakagawa:2017yiw}
Y.~O. Nakagawa, M.~Watanabe, S.~Sugiura and H.~Fujita, \emph{{Universality in
  volume-law entanglement of scrambled pure quantum states}},
  \href{https://doi.org/10.1038/s41467-018-03883-9}{\emph{Nature Commun.}
  {\bfseries 9} (2018) 1635}
  [\href{https://arxiv.org/abs/1703.02993}{{\ttfamily 1703.02993}}].

\bibitem{Vidmar:2017pak}
L.~Vidmar and M.~Rigol, \emph{{Entanglement Entropy of Eigenstates of Quantum
  Chaotic Hamiltonians}},
  \href{https://doi.org/10.1103/PhysRevLett.119.220603}{\emph{Phys. Rev. Lett.}
  {\bfseries 119} (2017) 220603}
  [\href{https://arxiv.org/abs/1708.08453}{{\ttfamily 1708.08453}}].

\bibitem{Murthy:2019qvb}
C.~Murthy and M.~Srednicki, \emph{{Structure of chaotic eigenstates and their
  entanglement entropy}},
  \href{https://doi.org/10.1103/PhysRevE.100.022131}{\emph{Phys. Rev. E}
  {\bfseries 100} (2019) 022131}
  [\href{https://arxiv.org/abs/1906.04295}{{\ttfamily 1906.04295}}].

\bibitem{Dong:2020iod}
X.~Dong and H.~Wang, \emph{{Enhanced corrections near holographic entanglement
  transitions: a chaotic case study}},
  \href{https://doi.org/10.1007/JHEP11(2020)007}{\emph{JHEP} {\bfseries 11}
  (2020) 007} [\href{https://arxiv.org/abs/2006.10051}{{\ttfamily
  2006.10051}}].

\bibitem{Ryu:2006bv}
S.~Ryu and T.~Takayanagi, \emph{{Holographic derivation of entanglement entropy
  from AdS/CFT}},
  \href{https://doi.org/10.1103/PhysRevLett.96.181602}{\emph{Phys. Rev. Lett.}
  {\bfseries 96} (2006) 181602}
  [\href{https://arxiv.org/abs/hep-th/0603001}{{\ttfamily hep-th/0603001}}].

\bibitem{Maloney:2016gsg}
A.~Maloney and S.~F. Ross, \emph{{Holography on Non-Orientable Surfaces}},
  \href{https://doi.org/10.1088/0264-9381/33/18/185006}{\emph{Class. Quant.
  Grav.} {\bfseries 33} (2016) 185006}
  [\href{https://arxiv.org/abs/1603.04426}{{\ttfamily 1603.04426}}].

\bibitem{Verlinde:2015qfa}
H.~Verlinde, \emph{{Poking Holes in AdS/CFT: Bulk Fields from Boundary
  States}},  \href{https://arxiv.org/abs/1505.05069}{{\ttfamily 1505.05069}}.

\bibitem{Nakayama:2016xvw}
Y.~Nakayama and H.~Ooguri, \emph{{Bulk Local States and Crosscaps in
  Holographic CFT}}, \href{https://doi.org/10.1007/JHEP10(2016)085}{\emph{JHEP}
  {\bfseries 10} (2016) 085}
  [\href{https://arxiv.org/abs/1605.00334}{{\ttfamily 1605.00334}}].

\bibitem{Minahan:1992ht}
J.~A. Minahan and A.~P. Polychronakos, \emph{{Integrable systems for particles
  with internal degrees of freedom}},
  \href{https://doi.org/10.1016/0370-2693(93)90395-X}{\emph{Phys. Lett. B}
  {\bfseries 302} (1993) 265}
  [\href{https://arxiv.org/abs/hep-th/9206046}{{\ttfamily hep-th/9206046}}].

\bibitem{Kazama:2014sxa}
Y.~Kazama, S.~Komatsu and T.~Nishimura, \emph{{Novel construction and the
  monodromy relation for three-point functions at weak coupling}},
  \href{https://doi.org/10.1007/JHEP01(2015)095}{\emph{JHEP} {\bfseries 01}
  (2015) 095} [\href{https://arxiv.org/abs/1410.8533}{{\ttfamily 1410.8533}}].

\bibitem{Jiang:2014cya}
Y.~Jiang, I.~Kostov, A.~Petrovskii and D.~Serban, \emph{{String Bits and the
  Spin Vertex}},
  \href{https://doi.org/10.1016/j.nuclphysb.2015.05.029}{\emph{Nucl. Phys. B}
  {\bfseries 897} (2015) 374}
  [\href{https://arxiv.org/abs/1410.8860}{{\ttfamily 1410.8860}}].

\bibitem{Jiang:2020sdw}
Y.~Jiang and B.~Pozsgay, \emph{{On exact overlaps in integrable spin chains}},
  \href{https://arxiv.org/abs/2002.12065}{{\ttfamily 2002.12065}}.

\bibitem{Vu:2019qxt}
D.-L. Vu, I.~Kostov and D.~Serban, \emph{{Boundary entropy of integrable
  perturbed SU (2)$_{k}$ WZNW}},
  \href{https://doi.org/10.1007/JHEP08(2019)154}{\emph{JHEP} {\bfseries 08}
  (2019) 154} [\href{https://arxiv.org/abs/1906.01909}{{\ttfamily
  1906.01909}}].

\bibitem{Kristjansen:2020vbe}
C.~Kristjansen, D.~M\"uller and K.~Zarembo, \emph{{Overlaps and fermionic
  dualities for integrable super spin chains}},
  \href{https://doi.org/10.1007/JHEP03(2021)100}{\emph{JHEP} {\bfseries 03}
  (2021) 100} [\href{https://arxiv.org/abs/2011.12192}{{\ttfamily
  2011.12192}}].

\bibitem{Kristjansen:2021xno}
C.~Kristjansen, D.~M\"uller and K.~Zarembo, \emph{{Duality relations for
  overlaps of integrable boundary states in AdS/dCFT}},
  \href{https://doi.org/10.1007/JHEP09(2021)004}{\emph{JHEP} {\bfseries 09}
  (2021) 004} [\href{https://arxiv.org/abs/2106.08116}{{\ttfamily
  2106.08116}}].

\bibitem{tsuchiya1998determinant}
O.~Tsuchiya, \emph{Determinant formula for the six-vertex model with reflecting
  end}, {\emph{Journal of Mathematical Physics} {\bfseries 39} (1998) 5946}.

\bibitem{Gombor:2021uxz}
T.~Gombor and B.~Pozsgay, \emph{{On factorized overlaps: Algebraic Bethe
  Ansatz, twists, and Separation of Variables}},
  \href{https://doi.org/10.1016/j.nuclphysb.2021.115390}{\emph{Nucl. Phys. B}
  {\bfseries 967} (2021) 115390}
  [\href{https://arxiv.org/abs/2101.10354}{{\ttfamily 2101.10354}}].

\bibitem{Gombor:2021hmj}
T.~Gombor, \emph{{On exact overlaps for $\mathfrak{gl}(N)$ symmetric spin
  chains}},  \href{https://arxiv.org/abs/2110.07960}{{\ttfamily 2110.07960}}.

\bibitem{Yao:2020dqx}
Y.~Yao and A.~Furusaki, \emph{{Parafermionization, bosonization, and critical
  parafermionic theories}},
  \href{https://doi.org/10.1007/JHEP04(2021)285}{\emph{JHEP} {\bfseries 04}
  (2021) 285} [\href{https://arxiv.org/abs/2012.07529}{{\ttfamily
  2012.07529}}].

\bibitem{Gaiotto:2020iye}
D.~Gaiotto and J.~Kulp, \emph{{Orbifold groupoids}},
  \href{https://doi.org/10.1007/JHEP02(2021)132}{\emph{JHEP} {\bfseries 02}
  (2021) 132} [\href{https://arxiv.org/abs/2008.05960}{{\ttfamily
  2008.05960}}].

\bibitem{Friedan:2003yc}
D.~Friedan and A.~Konechny, \emph{{On the boundary entropy of one-dimensional
  quantum systems at low temperature}},
  \href{https://doi.org/10.1103/PhysRevLett.93.030402}{\emph{Phys. Rev. Lett.}
  {\bfseries 93} (2004) 030402}
  [\href{https://arxiv.org/abs/hep-th/0312197}{{\ttfamily hep-th/0312197}}].

\bibitem{Casini:2016fgb}
H.~Casini, I.~S. Landea and G.~Torroba, \emph{{The g-theorem and quantum
  information theory}},
  \href{https://doi.org/10.1007/JHEP10(2016)140}{\emph{JHEP} {\bfseries 10}
  (2016) 140} [\href{https://arxiv.org/abs/1607.00390}{{\ttfamily
  1607.00390}}].

\bibitem{Caetano:2020dyp}
J.~a. Caetano and S.~Komatsu, \emph{{Functional equations and separation of
  variables for exact $g$-function}},
  \href{https://doi.org/10.1007/JHEP09(2020)180}{\emph{JHEP} {\bfseries 09}
  (2020) 180} [\href{https://arxiv.org/abs/2004.05071}{{\ttfamily
  2004.05071}}].

\bibitem{Cavaglia:2021mft}
A.~Cavagli\`a, N.~Gromov and F.~Levkovich-Maslyuk, \emph{{Separation of
  variables in AdS/CFT: functional approach for the fishnet CFT}},
  \href{https://doi.org/10.1007/JHEP06(2021)131}{\emph{JHEP} {\bfseries 06}
  (2021) 131} [\href{https://arxiv.org/abs/2103.15800}{{\ttfamily
  2103.15800}}].

\bibitem{Witten:1998xy}
E.~Witten, \emph{{Baryons and branes in anti-de Sitter space}},
  \href{https://doi.org/10.1088/1126-6708/1998/07/006}{\emph{JHEP} {\bfseries
  07} (1998) 006} [\href{https://arxiv.org/abs/hep-th/9805112}{{\ttfamily
  hep-th/9805112}}].

\bibitem{Caputa:2010ep}
P.~Caputa, C.~Kristjansen and K.~Zoubos, \emph{{On the spectral problem of N=4
  SYM with orthogonal or symplectic gauge group}},
  \href{https://doi.org/10.1007/JHEP10(2010)082}{\emph{JHEP} {\bfseries 10}
  (2010) 082} [\href{https://arxiv.org/abs/1005.2611}{{\ttfamily 1005.2611}}].

\bibitem{inprogress}
\emph{{In progress}}, .

\bibitem{Cordova:2018dbb}
C.~C\'ordova, G.~B. De~Luca and A.~Tomasiello, \emph{{Classical de Sitter
  Solutions of 10-Dimensional Supergravity}},
  \href{https://doi.org/10.1103/PhysRevLett.122.091601}{\emph{Phys. Rev. Lett.}
  {\bfseries 122} (2019) 091601}
  [\href{https://arxiv.org/abs/1812.04147}{{\ttfamily 1812.04147}}].

\bibitem{Cordova:2019cvf}
C.~C\'ordova, G.~B. De~Luca and A.~Tomasiello, \emph{{New de Sitter Solutions
  in Ten Dimensions and Orientifold Singularities}},
  \href{https://doi.org/10.1007/JHEP08(2020)093}{\emph{JHEP} {\bfseries 08}
  (2020) 093} [\href{https://arxiv.org/abs/1911.04498}{{\ttfamily
  1911.04498}}].

\bibitem{Cho:2015ega}
G.~Y. Cho, C.-T. Hsieh, T.~Morimoto and S.~Ryu, \emph{{Topological Phases
  Protected By Reflection Symmetry and Cross-cap States}},
  \href{https://doi.org/10.1103/PhysRevB.91.195142}{\emph{Phys. Rev. B}
  {\bfseries 91} (2015) 195142}
  [\href{https://arxiv.org/abs/1501.07285}{{\ttfamily 1501.07285}}].

\end{thebibliography}\endgroup
\end{document}